\documentclass[a4paper]{article}
\usepackage{graphicx}
\usepackage{amsfonts,amssymb,enumerate,epsf} 
\usepackage{authblk}
\usepackage{float}
 \usepackage[english]{babel}
\DeclareMathSizes{11}{9}{7}{5}
 \def\botcaption#1#2{\medskip\centerline{{\scshape #1.}\kern8pt
 {\rm #2}}\bigskip}
\def \ba {\begin{array}}
\def \ea {\end{array}}
\def \be {\begin{equation}}
\def \ee {\end{equation}}

\def\bal{\begin{aligned}}
\def\eal{\end{aligned}}

 \def \C {{\mathbb C}}
 
  \def \I {{\mathbb I}}

 \def \R {{\mathbb R}}
  
 \def \Z {{\mathbb Z}}

\def \bk{{\mathbf k}}
\def \bx{{\mathbf x}}
\def \bu{{\mathbf u}}
\def \bh{{\mathbf h}}
\def \bv{{\mathbf v}}
\def \bY{{\mathbf Y}}

 \def \cC {{\cal C}}

 \def \cL {{\cal L}}

 \def \cO {{\cal O}}

 \def \al {{\alpha}} 

 \def \ep {{\epsilon }}

\def \La {{\Lambda}}

 \def \p {{'}} 

\def \un {{(1)}}

\def \cst {{ \rm{const} }}
\def \zer {{ (0) }}
 \def \vet#1 {{\bf {#1}}}

\begin{document}
 
\title{On the  blow-up of some complex solutions of the 3-d Navier-Stokes Equations: Theoretical Predictions and Computer simulations} 
 \author{{\sc
 C. Boldrighini} $^{*}$ \\
{\it Istituto Nazionale di Alta Matematica (INdAM), GNFM, Unit\`a locale Universit\`a Roma Tre,  Largo S. Leonardo Murialdo, 1,  00146 Rome,
 Italy} \par
 $^{*}$ Corresponding author: boldrigh@mat.uniroma1.it 
  \\ \medskip
{\sc S. Frigio } and {\sc P. Maponi}\\
{\it Scuola di Scienze e Tecnologie, Universit\`a di Camerino (Italy)}  \\
 }
\maketitle
\bigskip
 \begin{abstract}
 {We consider some complex-valued solutions  of the Navier-Stokes equations in  $\R^{3}$ for which Li and Sinai proved a finite time blow-up.  We show that there are two types of  solutions, with different divergence rates, and report  results of computer simulations, which give a detailed picture of the blow-up for both types. They reveal in particular important features not, as yet, predicted by the theory, such as a concentration of the energy and the enstrophy  around a few singular points, while elsewhere the ``fluid'' remains quiet. 
 \par\medskip
 \hfill Dedicated to Yakov Grigorievich Sinai on the occasion of his 80-th anniversary}{}
%
  3-d Navier Stokes equations. Blow-up. Global regularity problem
\end{abstract}

 \section {Introduction}
 \label {S1}
 Since the beginning of the modern mathematical theory of the Navier-Stokes (NS) equations,  with  Jean Leray  in 1934 \cite{Leray},
   one of the main open questions is whether the solutions of the  initial value problem in $\R^{3}$   with smooth initial data and in absence of external forces can become singular in  a finite time (``global regularity problem'').     
  
 \par\smallskip
 Leray believed
         that the singular solutions   do exist,  and  are related to  turbulence. Although we now have a developed theory of turbulence which has no connection  to the NS singularities, it is clear  
        that  the existence of singular solutions could  provide a deeper understanding of physical phenomena  such as the fast concentration of energy in a small space region,  as we  see in  hurricanes.    We know in fact \cite{Seregin12} that a loss of smoothness  implies the divergence of the solution at some point of the physical space (and this is also shown by the Li-Sinai solutions).
       There is at present no effective model for such phenomena.  \par \smallskip
       Much work has been devoted to  the global regularity problem, both theoretical and by computer simulations.  
 An important   result was 
recently  obtained by T. Tao \cite{TT14}, who proved a finite-time blow-up  for modified NS equations which satisfy the energy conservation. The model is obtained by replacing the quadratic term in the NS equations  by a suitable average, and is related to the so-called ``dyadic'' model of Katz and Pavlovic, for which a finite-time blow-up can also be proved \cite{KP02}.  Other  theoretical results  have been obtained for suitable modifications of the equations (see  \cite{Ch08}, \cite{TT14} and references therein). 
\par\smallskip
       Li and Sinai \cite{LiSi08} consider a class of complex-valued solutions of    the classical initial value problem  in the whole space $\R^{3}$,  in absence  of boundary conditions and external forces
        \begin{equation} \label {1}  {\partial {\bf u}\over \partial t} + \sum_{j=1}^3 u_j{\partial\over \partial x_j} {\bf u} = \Delta \mathbf u - \nabla p, \qquad  {\bf x} = (x_1, x_2, x_3)\in \R^3 .  \end{equation} 
  $$ \nabla \cdot \mathbf u = 0, \qquad  \mathbf u(\cdot, 0) = \mathbf u_0. $$
 where   ${\mathbf u}  =(u_{1}, u_{2}, u_{3}): \R^3 \times [0, \infty) \to \C^3 $ is the complex ``velocity'' field, 
  $p$ is the (complex) pressure,   $\mathbf u_0$ is the initial data, and   the viscosity is $\nu=1$ (which is always possible by rescaling). If we   introduce the modified Fourier transform 
      \be\label{vu} \mathbf v(\mathbf k, t) = {i\over (2\pi)^{3}} \int_{\R^3} \mathbf u(\mathbf x, t) e^{i \langle \mathbf k, \mathbf  x\rangle } d{\mathbf x}, \qquad \mathbf k= (k_1, k_2, k_{3}) \in \R^3, \ee
    where $\langle \cdot, \cdot \rangle$ denotes the scalar product in $\R^3$, we are led, by a Duhamel formula, to  the   integral equation  $$ \mathbf v ({\bf k},t) = e^{-t {\bf k}^2} \mathbf v_{0}({\bk})   + $$
     \be\label{kequation} +  \int_0^t e^{-(t-s) {\mathbf k}^2} ds \int_{\R^3}\langle \mathbf v(\bk-\bk\p, s),\mathbf k \rangle    P_{\mathbf k}   \bv(\bk\p, s)   d \bk\p , \ee
    where   $ \bv_{0}$ is the transform of $\bu_{0}$, and $P_{\bk}$  is the solenoidal projector  expressing incompressibility     $$P_{\mathbf k} \mathbf v = \mathbf v - {\langle \mathbf v, \mathbf k\rangle \over {\bf k}^2} \mathbf k . $$
    
    Li and Sinai \cite{LiSi08} proved that there is a  class of real solutions of the integral equation (\ref{kequation}), corresponding to complex solutions of the NS equations (\ref{1}), which  become singular at a finite time (blow-up). Observe that we get a real NS solutions for odd initial data $\mathbf v_{0}$.\par\smallskip
    As we discuss below, the main  feature of such solutions, which is of great help both for the theoretical analysis and for computer simulations,  is that they have a rather simple structure in $\bk$-space: their support is concentrated in a thin cone around a fixed direction. \par\smallskip
 The total energy $E(t)$ and the totale enstrophy $S(t)$ (which measures the intensity of the vorticity field $\omega(\mathbf x,t) = \nabla \times \mathbf u(\mathbf x,t)$) 
    \begin{equation} \label{2} E(t) = {1\over 2} \int_{\R^{3}} |\mathbf u(\mathbf x,t)|^{2} d\mathbf x , \qquad S(t) =  \int_{\R^{3}} |\nabla \mathbf u(\mathbf x,t)|^{2} d \mathbf x =  \int_{\R^{3}} |\omega(\mathbf x,t)|^{2} d \mathbf x  .\end{equation}  
    are related by the law of conservation of energy 
     \begin{equation}\label{equality} E(t) + \int_{0}^{t} S(\tau) d\tau = E(0), \end{equation}
    which however  for complex solutions  is not coercive.  In fact, for the Li-Sinai solutions  both $E(t)$ and $S(t)$ 
    diverge as we approach the critical time.  \par\smallskip
    We recall that for real solutions (see, e.g., \cite{Temam}), if   $\mathbf u_{0}$ is such that $E(0)$ and $S(0)$ are finite, there is a unique regular solution in a maximal interval $t\in [0, T_{c})$, where the critical time $T_{c}$ depends on $\mathbf u_{0}$.
  Moreover  global regularity (i.e.,    $T_{c}  = +\infty $) holds  if $S(t)$ is finite for all $t>0$, a result which follows from an ``enstrophy inequality'' if $E(0)$ and $S(0)$ are small enough. 
   Therefore in the real case, if 
   $T_{c}$ is finite,   $S(t)$ diverges as $t\uparrow T_{c}$,  while $E(t)$ decreases by energy conservation,  implying a transfer of energy to the high $|\bk|$ modes as $t\uparrow T_{c}$. \par\smallskip 
    The main purpose of the present paper is to  obtain a better understanding of the relevant features of the complex Li-Sinai solutions, in particular of the peculiar ``mechanism '' by which   the high $\bk$-modes are enhanced, leading to the blowup. As we discuss in the next section, such mechanism is inherited by  the real solutions obtained by antisymmetrizing the initial data $\mathbf v_{0}$ of Li and Sinai,   although it is unclear whether  it is ``strong enough'' for a blow-up. In any case the  results of the present paper can be applied to the study of those solutions, a program that is already under way, shedding light on a new class of (real) fluid flows.
   
   \par\smallskip
     The computer simulations are of great help, as they reveal important properties which are not, so far, predicted by the theory. They indicate, in particular, point-wise convergence as $t\uparrow T_{c}$ of the solutions in the physical $\bx$-space, everywhere except at a few points (one point for solutions of type $I$ and two points for type $II$) around which the energy concentrates. \par\smallskip
     The computer simulations reported in the present paper required the implementation of a computational scheme for the integral equation (\ref{kequation}) which can be used for a class of complex and real solutions of the NS equation, including solutions that blow up.      
     \par\smallskip
   The plan of the paper is as follows.  In \S 2 we briefly report the Li-Sinai theory,  and 
 we   identify two different types of  solutions with different behavior near the critical time. We show that the solutions converge point-wise in $\bk$-space as $t\uparrow T_{c}$, and derive the divergence rates for $E(t)$ and $S(t)$  for the two types. In the following sections, \S 3,  and \S 4, we report the results of computer simulations.     Finally,     \S 5 gives some technical details on the computation, and \S 7 is devoted to  concluding remarks.

 \par \smallskip 
 We would like to remark that we simulate the solution of the integral equation (\ref{kequation}) for a $3$-d vector in $\R^{3}$,  with support going away to infinity, a task which requires   a grid of the order of $10^{9}$ points,  an unusual challenge even for modern supercomputers.  Reasonable results could be obtained only thanks to the relatively simple structure   of the Li-Sinai solutions  in $\bk$-space.

 \par\smallskip
 Computer simulations of a blowup for the integral equation (\ref{kequation}) were first  reported in \cite{AKh09}.
    However, due to computational limitations, their results only give  a qualitative description of the divergence of  energy and  enstrophy.\par\smallskip
  
Complex-valued solutions which blow up in a finite time with a similar behavior in $k$-space have been found also for the Burgers equations \cite{LiSi10} and other models \cite{LiSi10(2)}.  Computer simulations of blow-up solutions for the two-dimensional  Burgers equations,  which are much easier to handle, are reported in  \cite{BFM12}.

 \section {The Li-Sinai complex solutions and their divergence rates at the blow-up}
 \label {S2}
 
We begin by a   briefly description of the Li-Sinai theory, which is an excellent guideline to the understanding of the main features of the solutions. We refer the reader to the paper \cite{LiSi08} for the proofs and further details.
\par\smallskip

   As we mentioned above, we consider real solutions of the equation (\ref{kequation}), which in general correspond to  complex solutions $\bu(\bx,t)$. 
The initial data   $\bv_{0}$ are  localized around a point $\bk^{\zer}$, at a certain distance from the origin.  In our simulations we   took $\bk^{\zer}= (0, 0, a)$ with  $5\leq a  \leq 25$, and the  support of $\mathbf v_{0}$  in a circle with center $\bk^{\zer}$ and radius $r<a$.   \par\smallskip
 Multiplying the initial data  $\bv_{0}$ by  a positive parameter   $A$ and iterating the Duhamel formula we can write the solution  as a power series 
 \begin{equation}\label{serie} \bv_{A}(\bk, t) =  A   e^{- t {\bk}^2} \bv_{0}(\bk) + \int_0^t e^{-{\mathbf k}^2(t-s) } \sum_{p=2}^\infty A^p \mathbf g^{(p)}(\mathbf k, s) ds. \end{equation}
     Substituting into the equation,   we see that  the functions   $\mathbf g^{(p)}(\mathbf k, s)$ satisfy a recursive relation which reminds the BBGKY hierarchy of Statistical Physics. Setting $\mathbf g^\un(\mathbf k, s) = e^{- s {\bk}^2} \bv_{0}(\bk)$      and
 $$  \mathbf g^{(2)}(\mathbf k, s) =   \int_{\R^3} \left \langle \bv_{0}(\bk-\bk\p),\bk \right \rangle   
  P_{\bk} \bv_{0}(\bk\p)   e^{- s(\bk - \bk\p)^2 - s (\bk\p)^2} d {\bk}\p ,$$
 we find for $p>2$ the recursive relation
 
 $$    \mathbf g^{(p)}(\mathbf k, s) =  $$
 $$ =  \int_0^s ds_2 
    \int_{\R^3}  \left \langle \bv_{0}(\bk-\bk\p),\bk \right \rangle    P_{\bk}  \mathbf g^{(p-1)}(\bk^\p, s_{2}) e^{- s(\bk - \bk\p)^2 - (s-s_{2}) {(\bk}^\p)^2} d {\bk}^\p  + $$
 \be\label{0}+  \sum_{p_1 + p_2 = p\atop p_1, p_2 >1} \int_0^s ds_1 \int_0^s ds_2 
   \int_{\R^3}\left  \langle \mathbf g^{(p_1)}(\mathbf k - \mathbf k\p, s_1), \bk \right \rangle \cdot $$ $$\cdot   P_{\bk}\mathbf g^{(p_2)} (\bk\p, s_2) e^{-  (s-s_1) (\bk - \bk\p)^2 - (s-s_2) ( \bk\p)^2} d\mathbf k\p  + \ee
 $$ +   \int_0^s ds_1 \int_{\R^3}  \left  \langle \mathbf g^{(p-1)}(\bk - \bk\p, s_1), \bk \right  \rangle   P_{\bk} \bv_0(\bk\p) e^{-  (s-s_1) (\mathbf k - \mathbf k\p)^2 - s ( \bk\p)^2} d\bk\p.$$

\par\smallskip

If $C= \rm{supp} \; \bv_{0}$, then, by iteration of the convolution, the support of $\mathbf g^{(p)}$ will be $\underbrace {C + \ldots + C}_{ p \;  times}$.  As $C$ is around $\bk^{\zer} = (0,0, a)$,  the support of the solution extends along the $k_{3}$-axis.  \par \smallskip 
 
 By analogy with the  theory of probability, where  convolution corresponds to the distribution of a sum of random variables, 
 we know that for
 large $p$ the  support of $\mathbf g^{(p)}$ is  around  $p \bk^{\zer}$, in a region with transversal dimensions of the order $\sqrt {p}$. 
Moreover if $p$  is large  the terms of the sum for which $\max \{ p_{1}, p_{2}\} \leq p^{1\over 2}$  can be neglected, and the Gaussian densities   give a significant contribution to the integrals only for $s_{1}, s_{2}$ near the endpoint $s$. Therefore we introduce the  new variables and  functions 
\be\label{nvf} \bk = p \bk^{\zer} + \sqrt {p}   \mathbf Y, \quad \mathbf h^{(p)}(\mathbf Y, s) = \mathbf g^{(p)}( p \bk^{\zer} + \sqrt {p }  \bY, s), \quad
 s_{j} = s  \left ( 1 - {\theta_{j}\over p^{2}_{j}} \right ), \quad j=1,2 . \ee
 \par \smallskip
 
Integrating over $\theta_{j}, j=1,2$ and  setting $\gamma = {p_{1}\over p}$ we get 
\be\label{4}  \mathbf h^{(p)} (\bY, s) =   {p^{5\over 2}}  \sum_{p_{1}+p_{2}= p\atop p_{1},p_{2} > \sqrt p}   {1\over p_{1}^{2} p_{2}^{2}} \int_{\R^{3}} P_{ \mathbf e_{3}+ {\bY \over \sqrt {p}}}   \mathbf h^{(p_{2})} \left ( {\bY\p\over \sqrt {1-\gamma}}, s \right )\cdot   \ee 
$$\cdot  \left \langle \mathbf h^{(p_{1})}   \left ( {\bY - \bY\p\over \sqrt \gamma}, s \right ), \mathbf e_{3}+ {\bY \over \sqrt {p}} \right \rangle  d\bY\p \; \left (1 + o(1) \right ),$$
where $\mathbf e_{3}= (0, 0, 1)$.  We write the components of $\bh^{(p)}(\bY, s)$ in the form 
  \be\label{5}\bh^{(p)} (\bY, s)= \left ( H^{(p)}_{1}(\bY, s),  H^{(p)}_{2}(\bY, s), {F^{(p)}(\bY, s)\over \sqrt {p} \; a}  \right ),\ee
and as, by incompressibility, $\bh^{(p)}$  is orthogonal to $\bk = ( \sqrt {p}  Y_{1}, \sqrt {p}  Y_{2}, pa + \sqrt {p}  Y_{3} )$,   we get
$$ Y_{1} H^{(p)}_{1}(\bY, s) + Y_{2} H^{(p)}_{2}(\bY, s) + F^{(p)}(\bY, s) = \cO(p^{-{1\over 2}}a^{-1}) ,$$ 
 which shows that $F^{(p)}(\bY, s)$ is of finite order. 
Therefore 
  $\mathbf h^{(p)} (\bY, s)$,  is essentially transversal to the $k_{3}$-axis, and as $p\to\infty$,
    $P_{ \mathbf e_{3}+ {\bY \over \sqrt {p}}}   \mathbf h^{(p_{2})} \to  \mathbf h^{(p_{2})} $,  i.e., the solenoidal   projector in (\ref{4}) tends to      the identity.   
  \par \smallskip 
   The fundamental {\it Ansatz} is that for some set of initial data $\mathbf v_{0}$, when  $p$ is large and  $s$ in some interval of time,  the recursive relation (\ref{4}) has an approximate  solution which is asymptotically of the form
 \begin{equation}\label{ansatz} \mathbf h^{(p)}(\bY, s) \; =  \;    p \; (\Lambda(s))^{p}  \prod_{j=1}^{3}  g_{\sigma_{j}}(Y_{j})  \left ( \mathbf H(\bY) + \mathbf \delta^{(p)}(\bY, s) \right ) .\end{equation}
 Here   $\La(s)$ is a positive  function, which will be discussed below,    $g_{\sigma}(x) ={ e^{- {x^{2}\over 2 \sigma}}\over  \sqrt {2\pi\sigma} } $ denotes the centered Gaussian density on $\R$, $\sigma_{1}, \sigma_{2}, \sigma_{3}$ are positive constants,   $\mathbf H$ is a   vector function independent of time,  orthogonal to   $\mathbf e_{3}$,  and depending only on $Y_{1}, Y_{2}$,
 $$\mathbf H(\bY) = \left (H_{1}(\bY),   H_{2}(\bY), 0 \right ),$$ 
 and the remainder \begin{equation} \label {6}\delta^{(p)}( \bY, s) =
  \left (  \delta_{1}^{(p)}(\bY, s),    \delta_{2}^{(p)}(\bY, s),   \delta_{3}^{(p)}(\bY, s)\right ) \end{equation}
  is    such that $\delta^{(p)}(\bY, s) \to 0$ as $p\to \infty$.  
  \par\smallskip
 It is important to observe that by the  {\it Ansatz} (\ref{ansatz}), the function  $\mathbf h^{(p)}(\bY, s)$ is proportional to a product of Gaussian functions, and the time dependence of its leading term is determined by the  function $\Lambda(s)$.

 \par\smallskip
In view of possible rescalings 
it is not restrictive to set  $\sigma_{i}=1, i=1, 2, 3$.  Inserting (\ref{ansatz}) into (\ref{4}), treating $\gamma$ as a continuous variable, neglecting the remainders, and integrating over $Y_{3}$,  one can see that  $\mathbf H(\bY)$ is a solution of the  integral fixed point equation
 \be\label{7} g^{(2)}_{1}(\bY)  \mathbf H(\bY) =\int_{0}^{1} d\gamma\int_{\R^{2}} g^{(2)}_{\gamma}(\bY-\bY\p) g^{(2)}_{1-\gamma}(\bY\p)   \cL(\mathbf H; \gamma, \bY, \bY\p)   \mathbf H\left ( {\bY\p\over \sqrt{1-\gamma}}\right ) d\bY\p\ee
 where, by abuse of notation, $\mathbf H = (H_{1}, H_{2})$ and $\bY = (Y_{1}, Y_{2})$ are in $\R^{2}$,    $ g^{(2)}_{\sigma}(\bY) = {e^{- {Y_{1}^{2}+Y_{2}^{2}\over 2 \sigma} }\over 2\pi \sigma}$,  and   $$  \cL(\mathbf H; \gamma, \bY, \bY\p)  = - (1-\gamma)^{3\over 2} \left \langle {\bY - \bY\p\over \sqrt \gamma},   \mathbf H \left ({\bY - \bY\p\over \sqrt \gamma} \right ) \right \rangle+$$
 $$+   \gamma^{1\over 2} (1-\gamma) \left \langle { \bY\p\over \sqrt {1-\gamma}},   \mathbf H \left ({\bY\p\over \sqrt {1-\gamma}} \right ) \right \rangle.$$
  \par \smallskip

  The solutions, or ``fixed points'',  of the functional equation (\ref{7})  are found by expanding $\mathbf H$ in Hermite polynomials ${\rm He}_{k}, k=0,1,\ldots$, which are orthogonal with respect to the standard Gaussian,
 \be\label{hermite} H_{j}(\bY) = \sum_{m_{1}, m_{2}=0}^{\infty} \ell^{(j)}_{m_{1}m_{2}} {\rm He}_{m_{1}}(Y_{1})  \; {\rm He}_{m_{2}}(Y_{2}), \qquad j=1,2 .\ee
There are infinitely many fixed points, and there is a  class $\cC$ of  them such that  for  $\mathbf H \in \cC$ there is an open set of initial data $\bv_{0}$ for which  the  solution 
satisfies  the {\it Ansatz} (\ref{ansatz}) for $s\in S = [s_{-}, s_{+}]$, where $S$ is a non-empty  time interval. 
The open set is constructed by linearized stability analysis, and the proofs are based  on the renormalization group method.\par\smallskip
Following Li and Sinai we consider the behavior of the solution near the blow-up time for a fixed point which is proportional to the radial vector
\be\label{accazero}\mathbf H^{\zer} = c\;  \mathbf Y_{\perp}, \qquad \mathbf Y_{\perp} =  (Y_{1}, Y_{2}, 0 ), \ee
where $c>0$ is a suitable constant determined by the fixed point analysis, and from now on we use the pedex $\perp$ to denote vector components perpendicular to the $k_{3}$-axis. \par\smallskip

 A delicate analysis (see  \cite{LiSi08} and references therein) shows that the function $\Lambda(s)$ is differentiable and strictly increasing.\par \smallskip
 
 Observe that the recursive relation (\ref{4}) is unchanged if we replace $\mathbf h^{(p)}$ with $(-1)^{p} \mathbf h^{(p)}$, and the fixed point equation (\ref{7}) is also unchanged. Therefore in the fundamental  {\it Ansatz} (\ref{ansatz}) we can replace the term $(\Lambda(s))^{p}$  by $(-1)^{p} (\Lambda(s))^{p}$.\par \smallskip 
 We  get two types of solutions, those for which $h^{(p)}$ is given by (\ref{ansatz}), which will be called solutions of type $I$, and  those for which  $h^{(p)}$ is replaced by $(-1)^{p} h^{(p)}$,  which will be called solutions of type $II$.  Observe that  if the initial data $\bv_{0}$ is chosen according to the prescriptions in \cite{LiSi08}  and leads to a solution of type $I$ the fixed point $\mathbf H^{\zer}$, then the initial data $-\bv_{0}$ lead to a solution of type $II$ with the same fixed point.   As we shall see below,  the two types behave rather differently.
 
 \par\smallskip
 
  We now set  $A={1\over \Lambda(\tau)}$, for $\tau \in S$, and consider the solution (\ref{serie}) for $t< \tau$.    Taking into account (\ref{accazero}),  setting $\sigma_{I}(p) = 1$, $\sigma_{II}(p) = (-1)^{p}$, $p=1, 2, \ldots$, 
  the tail of the series in equation (\ref{serie}) can be replaced for both types  by its asymptotics
    \be\label {8}    \bv^{(p_{0}, \al)}(\bk, t)  =    C \;  
     \sum_{p=p_{0}}^\infty p\; \sigma_{\al}(p)\;  g^{(3)}(\bY^{(p)}) \; 
     \mathbf \bY_{\perp}^{(p)}  \int_{0}^{t}  e^{-{\mathbf k}^2(t-s) } \left ( {\La(s)\over \La(\tau)} \right )^{p} ds, \qquad \al = I, II  .\ee
Here we write for clarity  $\bY^{(p)} = {\bk - p \bk^{\zer}\over \sqrt p}$,   $p_{0}$ is large enough for the asymptotic behavior to set in, $C$ is a constant, and    $g^{(3)}$   is the three-dimensional standard Gaussian density. The explicit asymptotics (\ref{8}) shows that the main support of the solution extends along the direction $\mathbf k^{\zer}$, i.e., along the positive $k_{3}$-axis in a thin cone of transversal diameter proportional to $\sqrt {k_{3}}$.  \par\smallskip
\par\smallskip
As  $\Lambda$ is strictly increasing,  ${\Lambda(s) \over \Lambda(\tau)} < {\Lambda(t) \over \Lambda(\tau)}  <1$, and the terms of the series (\ref{8}) fall off exponentially fast in $p$, with an exponential rate  which vanishes as $t\uparrow \tau$. Therefore the critical time is $T_{c}=\tau$, and, as the single terms of the series (\ref{serie}) tend to a finite limit as $t\uparrow \tau$, any   divergence  is due to the tail.  \par\smallskip

We now pass to a discussion of the main properties of the blow-up,  based on the explicit asymptotic expression of the tail  series (\ref{8}). Although the results that follow are new, we will only give a sketch of the proofs. Full proofs are straighforward, but would require taking care of the corrections, which at some points is a rather cumbersome procedure.  In what follows we may denote by $\cst$ different constants.\par 

 \par\smallskip

As a first step we give a more convenient representation of the tail series (\ref{8}), which is valid when $t$ is close to $\tau$. (Observe that  it is not restrictive to assume that    the initial time is close   to $\tau$.) Since  $\Lambda(s)$ is differentiable, we have,  as $s\uparrow \tau$
\be\label{smooth} \ln {\Lambda(s)\over \Lambda(\tau)} = - \kappa (\tau-s) (1 + r(t-s)), \qquad \kappa = {\Lambda^{\prime}(\tau)\over \Lambda(\tau)}\ee
where $\kappa >0$ and   $r(s) \to 0$ as $s \to 0$. Neglecting the remainder,  and integrating  
  over $s$,   if $p_{0}$ is large, we get as $t\uparrow \tau$, for $\al = I, II$,
   \be\label{mainp} \bv^{(p_{0}, \al)}(\bk, t)  \sim \cst \;  \tilde \bv^{(p_{0},\al)}(\bk, t), \qquad \tilde \bv^{(p_{0}, \al)}(\bk, t)   =       \sum_{p\geq p_{0}} p \; \sigma_{\al}(p) \;  {e^{-\kappa p (\tau-t)}\over \bk^{2}+ \kappa p}    g^{(3)}(\bY^{(p)})\;    \mathbf Y_{\perp}^{(p)} .  \ee 
\par\smallskip
 Let now  $\bv^{(\al)}_{A(\tau)}(\bk, t)$, $\al = I, II$,  be the solution expressed by the series (\ref{serie}),  where $A=A(\tau)={1\over \La(\tau)}$, with $\tau >0$ and the initial data $\pm \bv_{0}$ leading to the fixed point $\mathbf H^{\zer}$ chosen as explained above (and in more detail in \cite{LiSi08}).
 We  first show  that, as  $t\uparrow \tau$,  there is a point-wise limit, i.e., for any fixed $\bk$ 
  \be\label{pointwise} \lim_{t\uparrow \tau} \bv^{(\al)}_{A(\tau)}(\bk, t) = \widehat \bv^{(\al)}_{A(\tau)}(\bk), \qquad \al = I,II. \ee
 \par\smallskip

 In fact the series (\ref{mainp}) is bounded in absolute value by 
$$| \bk_{\perp}|  \sum_{p=p_{0}}^{\infty}{p^{1\over 2}\over \bk^{2}+ \kappa p} \; g^{(3)}\left ({\bk - p \bk_{0}\over \sqrt p} \right ), $$and the terms for $p> \bk^{2}$ are bounded by $c\; p^{-{1\over 2 }}  \;e^{-{p a^{2}\over 4}}$, 
where the constant  $c$ depends on $\bk$. Hence the series (\ref{mainp}) converges absolutes, and so does the original series (\ref{serie}).
 
 \par\smallskip

Another point which was not considered in the previous literature is that of the rate of divergence of the totale energy $E(t)$ and the totale enstrophy $S(t)$.  We will now show that the divergence rates  are different  for solutions of the two types. More precisely, as $t\uparrow \tau$,  we have 
 \be\label{divergenza} E(t) \sim {C_{E}\over (\tau-t)^{\beta_\al}}, \qquad  S(t) \sim {C_{S}\over (\tau-t)^{\beta_\al+2}}, \qquad\al =I, II \ee
 where $\beta_{I} = 1$,  $\beta_{II}= {1\over 2}$ and $C^{(\al)}_{E}, C^{(\al)}_{S}$ are constants depending on the initial data.\par\smallskip
By what we said above, the divergence rate is determined by the asymptotic tail $\tilde \bv^{(p_{0}, \al}$ in (\ref{mainp}). 
 For solutions of type $I$,  as $\langle \bY^{(p_{1})}_{\perp},  \bY^{(p_{2})}_{\perp} \rangle = {k_{1}^{2}+ k_{2}^{2}\over \sqrt{p_{1}p_{2}}}$,  and setting $R_{p}(\bk) = \sqrt p { g^{(3)}(\bY^{(p)})\over |\bk|^{2}+ \kappa p}$ we have

   \be\label {vmsq} \left | \tilde \bv^{(p_{0}, I)}(\bk, t) \right |^{2} = \sum_{p=p_{0}}^{\infty} e^{- 2 p \kappa (\tau-t)} |\bk_{\perp}|^{2} \left [ R^{2}_{p}(\bk) + 2 \sum_{j=1}^{\infty}e^{-j\kappa(\tau-t)} R_{p}(\bk)R_{p+j}(\bk) \right ] .\ee
   The product $R_{p}(\bk)R_{p+j}(\bk)$ has a factor $\exp\{-{1\over 2}( |\bY^{(p+j)}|^{2} + |\bY^{(p)}|^{2})\}$, and writing   $\bY^{(p+j)}$, $j=0, 1, \ldots$,   in terms of  $\bY:=\bY^{(p)}$ we find 
   \be\label{nfinally} |\bY|^{2}+ \left |\bY^{(p+j)}\right |^{2} = {j^{2} a^{2}\over 2p +j} + {2p+j\over p+j}  \left |\tilde \bY^{p,j}\right |^{2 }, \qquad \tilde \bY^{p,j}=  \bY - \sqrt p {j \;\bk^{\zer} \over 2 p + j}  ,\ee
so that for  the integrals of the terms of the series (\ref{vmsq}) we get
    \be\label{defipj} I_{p,j} : =  \int_{\R^{3}} |\bk_{\perp}|^{2} R_{p}(\bk) R_{p+j}(\bk) d\bk =  {\sqrt {p (p+j)}\over (2\pi)^{3}} \int_{\R^{3}} |\bk_{\perp}|^{2}  e^{- {j^{2}a^{2}\over 2p +1}}{ \exp \{-{2p +j\over 2(p+j)} (|\tilde \bY^{p,j}|^{2}  \}\over (|\bk|^{2} + \kappa p) \; (|\bk|^{2} + \kappa (p+j)) }d\bk . \ee
  Let  $j_{0}(p) : = [p^{{1\over 2} + \ep}]$, $\ep \in (0,{1\over 2})$, where $[\cdot]$ is  the integer part.  Due to the factor $\exp \{- {1\over 2} {j^{2}a^{2}\over 2p + j}\}$ we have 
   \be\label{resto1} \sum_{j> j_{0}(p)} I_{p,j} = o(e^{- c_{1} p^{2\ep}}), \qquad c_{1}>0.\ee
   \par\smallskip 
Observe that  due to the Gaussian factor, the 
integral (\ref{defipj}) in the region $|\tilde \bY^{p,j}| > p^{\ep\over 2}$ falls off faster than any inverse power of $p$, and if  $j\leq j_{0}(p)$ and $|\tilde \bY^{p,j}| \leq  p^{\ep\over 2}$, setting  $s_{j} = {j\over \sqrt p}$, we have
\be\label{nfinally1}  |\bk|^{2} = \left | \sqrt p \bY + p\bk^{\zer}\right |^{2} = p^{2} \left | {1\over\sqrt p} (\tilde \bY^{p,j}+ { p\over 2p+j} s_{j}\bk^{\zer}) + \bk^{\zer}\right |^{2} \sim p^{2} |\bk^{\zer}|^{2} =   p^{2} a^{2}.\ee 
 For  $p$ large   $\tilde \bY^{p,j} \sim \tilde {\bY}^{(s_{j})} := \bY - { s_{j}\over 2 } \bk_{0}$, and   observing that  
$|\bk_{\perp}|^{2} = p |\tilde \bY_{\perp}^{(s_{j})}|^{2}$, and changing the integration variable     to $\tilde \bY^{(s_{j})}$,  we get the following asymptotics, uniformly in  $j= 0,1, \ldots, j_{0}(q)$,  
\be\label{offdiag2} I_{p,j} \sim   p^{-{1\over 2}}{e^{- {s^{2}_{j} a^{2}\over 4}}\over (2\pi)^{3}  a^{4}} \int_{\R^{3}} |\tilde \bY^{(s_{j})}_{\perp}|^{2} \exp\{ - |\tilde \bY^{(s_{j})}|^{2}\} d\tilde \bY^{(s_{j})} =  {1\over \sqrt p}{e^{- {s^{2}_{j} a^{2}\over 4}}\over (4\pi)^{3\over 2}  a^{4}}   .     \ee 
Finally, as $s_{j}\leq s_{j_{0}(q)}\leq p^{\ep}$, we have, as  $p\to \infty$,
\be\label{agg} \sum_{j=1}^{\infty }  e^{- {s^{2}_{j} a^{2}\over 4}} e^{-{s_{j}\over \sqrt p}\kappa (\tau-t)} {1\over \sqrt p} \;  \sim \sum_{j=1}^{\infty }  e^{- {s^{2}_{j} a^{2}\over 4}}   {1\over \sqrt p}  \to \; \int_{0}^{\infty} e^{- {s^{2} a^{2}\over 4}} ds = {\sqrt \pi\over a}.\ee
Therefore the sum $\sum_{j=1}^{j_{0}(p)} e^{-\kappa j(\tau-t)} I_{p,j}$ tends to a constant for large $p$,  and gives the main contribution to  the energy. In conclusion for  $t\uparrow \tau$ we have 
 \be\label {asympen1} E(t) \sim \cst  \int_{\R^{3}} | \tilde \bv^{(p_{0}, I)}(\bk, t)|^{2} d \bk = \cst \sum_{p\geq p_{0}}  e^{-2 \kappa p(t-\tau)} \left [ I_{p,0} + 2 \sum_{j=1}^{\infty} e^{-j \kappa (\tau-t)} I_{p,j} \right ] \sim \ee
 $$\sim \cst  \sum_{p\geq p_{0}} e^{- 2\kappa p(\tau-t)} \sim {\cst \over \tau-t}, $$
which  proves the first asymptotics (\ref{divergenza}) for solutions of type I.
   \par\smallskip
   For  $ \tilde \bv^{(p_{0},II)}$ the series in (\ref{mainp}) has alternating signs, and 
      (\ref{vmsq}) is replaced by  
     \be\label {vmsqII} \left | \tilde \bv^{(p_{0},II)}(\bk, t) \right |^{2} = \sum_{p=p_{0}}^{\infty} e^{- 2 p \kappa (\tau-t)} |\bk_{\perp}|^{2} \left [ R^{2}_{p}(\bk) + 2 \sum_{j=1}^{\infty}e^{-j\kappa(\tau-t)} R_{p}(\bk)   \tilde R_{p+2j-1}(\bk,t)   \right ] ,\ee
    $$ \tilde R_{p+\ell-1}(\bk, t) =  -R_{p+\ell-1}(\bk)+ e^{-\kappa(\tau-t) } R_{p+ \ell}(\bk), \qquad \ell =1, 2, \ldots\; .$$

    The contribution of the diagonal terms to $E(t)$ is the same as for the solutions of type $I$, and, taking into account the asymptotics (\ref{offdiag2}),   we have as $t\uparrow \tau$, 
   \be\label{IIdiagonal} \sum_{p\geq p_{0}} e^{-2 \kappa p(t-\tau)}  I_{p,0} \; \sim \; \cst  \sum_{p\geq p_{0}} {e^{-2p\kappa(\tau-t)} \over \sqrt p} \sim    {\cst  \over \sqrt {\tau - t}}.  \ee  
   \par\smallskip
   For the off-diagonal terms, observe   that, as for type $I$,    the  sum for  $j > j_{0}(p)$ gives a negligible contribution for large $p$.  Furthermore, we have
   \be\label{terre}   \tilde R_{p+\ell-1}(\bk, t)  =R_{p+\ell-1}(\bk) \left [  F_{p,\ell}(\bk)-1 \right ] + \cO(\tau-t) R_{p+\ell}(\bk), \ee
   $$F_{p,\ell}(\bk) = {R_{p+\ell}(\bk) \over R_{p+\ell-1}(\bk)}  =    \sqrt {p+\ell \over p+\ell-1} {|\bk|^{2} + \kappa (p+\ell-1) \over |\bk|^{2} + \kappa(p+\ell )} \exp\left \{ {1\over 2} \left [|\bY^{(p+\ell-1)}|^{2} - |\bY^{(p+\ell)}|^{2} \right ]\right \}.$$
   Writing $\bY^{p+\ell-1}, \bY^{p+\ell}$ in terms of $\bY^{p}=: \bY$ we see that
   \be\label {diff}  | \bY^{(p+\ell)}  |^{2}-  | \bY^{(p+\ell-1)}|^{2} = - {p\over (p+\ell) (p+\ell-1)} |\bY -   s_{\ell-1}\bk^{\zer}|^{2} + {a^{2}\over p+\ell} - 2 a\sqrt p {Y_{3}- a s_{\ell-1}  \over p+\ell} .\ee
   \par\smallskip 
   For $j\leq j_{0}(q)$, taking into account (\ref{terre}), where we neglect the term of order $\cO(\tau-t)$,  we need to compute the integrals
   \be\label {jayp} J_{p,j} =   \int_{\R^{3}}|\bk_{\perp}|^{2} R_{p}(\bk)  R_{p+2j-1}(\bk)  \left [ F_{p,2j}(\bk)-1 \right] d\bk   ,\ee
   where  the exponential in the definition of $F_{p, 2j}$ is written in the form (\ref{diff}). 
   We then change the integration variable to $\tilde \bY^{(s_{2j-1})}$, and observe that  in the region $|\tilde \bY^{(s_{2j-1})}|\leq p^{\ep\over 2}$, neglecting terms of the order $\cO(p^{-1+2\ep})$, we have
   $$F_{p, 2j} \sim \exp \left \{ {a  \sqrt p{Y_{3}- a s_{2j-1}\over  p+ 2j}} \right \} \sim 1 + {a\over \sqrt p} \left [ \tilde Y_{3}^{(s_{2j-1)}} + {a s_{2j-1}\over 2}   \right ].$$
   The term $\tilde Y_{3}^{(s_{2j-1})}$ given no contribution to the integral by parity, so that  for large $p$
  $ J_{p,j}\sim {a^{2}\over 2 \sqrt p} s_{2j-1} I_{p,2j-1} $,  where $I_{p, 2j-1}$ is given by (\ref{offdiag2}). Summing over $j$, as in (\ref{agg}), we find
   \be\label{svolta}  \sum_{j=1}^{j_{0}(q)}  J_{p,j} e^{-\kappa (2j-1) (\tau-t)} \sim {\cst\over \sqrt p}  {a^{2}\over 2 \sqrt p}\sum_{j=1}^{j_{0}(q)} s_{2j-1} {e^{- {s^{2}_{2j-1} a^{2}\over 4}}}\sim {\cst\over \sqrt p}, \ee
   where the constants are positive. 
   Proceeding as for the diagonal part (\ref{IIdiagonal}) we see that the off-diagonal terms give a contribution which is asymptotically of order $(\tau-t)^{-{1\over 2}}$. \par \smallskip
   
   The first relation (\ref{divergenza}) is proved.\par\smallskip

   For the second relation (\ref{divergenza}) it is enough to observe that the enstrophy density in $\bk$-space is proportional to the energy density multiplied by a factor $|\bk|^{2}$, which, taking into account the asymptotics (\ref{nfinally1}) produces an additional factor $p^{2}$ in the analogues of the integrals $I_{p,j}$. \par\smallskip
   As a final remark for this section, we would like to point out that, as shown by  the series representation (\ref{serie}), the Li-Sinai solutions describe a peculiar mechanism of enhancing the high $\bk$-modes, which is due to convolutions of modulated Gaussian terms with centers on the $k_{3}$-axis at the points $k_{3} = pa$, $p\in \Z_{+}$. \par \smallskip  It is easy to see that the a similar mechanism is going to work also for the real solutions obtained by antisymmetrizing the Li-Sinai initial data, although it is not clear how effective it will be, because the Gaussian centers will be a sum of positive and negative terms and tend to be closer to the origin.

     \section {Computer Simulations: the Li-Sinai solutions in $\bk$-space}
 \label {S3}
   We simulate the integral equation (\ref{kequation})  on a regular mesh in $\bk$-space contained in the region $ R = [-127, 127] \times [-127, 127] \times [-19, L]$. The parameter $L$, as we describe below, is of critical importance, and for the simulations reported below it takes the values  $2028, 2528, 3028$. In what follows we will only indicate the simulation range of the longitudinal variable $k_{3}$. More details are given in \S 6.
 
    \par\smallskip
 
The  initial data $\bv_{0}$  are chosen according to the prescriptions  in \cite{LiSi08}  (\S 7, formula (39)), with support concentrated around the point $\mathbf k^{(0)}=(0, 0, a)$, $a>0$, and  leading to the fixed point  (\ref{accazero}).    
A full screening for the ``best'' cases with a large random choice of the initial parameters, as in our  paper on the 2-d Burgers equations \cite{BFM12}, was not possible because for the 3-d NS equations   it takes too much computer time. We only considered    about a hundred cases and followed up the most promising ones.
\par\smallskip

For all results reported in this paper the initial data are of the following form
 \be\label{iniziali} \bv_{0}^{\pm}(\bk) = \pm\;   C\;  \bar{\bv}_{0}(\bk)\;  \I_{D}(\bk-\bk^{\zer}), \qquad  \bar{\bv}_{0}(\bk) =  \left ( k_{1}, k_{2}, - {k_{1}^{2}+ k_{2}^{2} \over k_{3} }\right ) {e^{-{(\bk - \bk^{\zer})^{2}\over 2}}\over (2\pi)^{3\over 2}  },\ee 
where $a=20$,  $\I_{D}$ is the indicator function of the support $D= \{ \bk: |\bk |\leq 17\}$, and the positive constant $C$ controls the initial energy and enstrophy.  
The initial data  $\bv_{0}^{+}$ lead to solutions of type $I$, and the initial data $\bv_{0}^{-}$ to  solutions with oscillating sign of type $II$. 

\par \smallskip

   We studied in more detail the solutions of type II, as their behavior is  more similar to that of the real solutions obtained by antisymmetrizing the initial data, which is an  object of our present research.

  \par\smallskip
In describing  the blow-up  an important role is played, due to the structure of the solutions,  by the marginal densities for energy and enstrophy  along the  $k_{3}$ axis in $\bk$-space
 \be\label{3k} E_{3}(k_{3}, t) = \int_{\R\times \R} d k_{1} dk_{2} e(\mathbf k, t) \qquad S_{3}(k_{3}, t) = \int_{\R\times \R} d k_{1} dk_{2} s(\mathbf k, t) ,\ee
$$ e(\bk, t) = {1\over 2} |\bv(\bk, t)|^{2}, \qquad s(\bk, t) =  |\bk|^{2} |\bv(\bk, t)|^{2}.$$
 By our definition of the transform (\ref{vu}) the total energy and enstrophy in (\ref{2}) are given by
 $$E(t) = {(2\pi)^{3}\over 2} \int_{\R^{3}} e(\mathbf k, t) d\bk , \qquad S(t) = (2\pi)^{3} \int_{\R^{3}} s(\mathbf k, t) d\bk.$$
The marginal densities along the third axis  in $\bx$-space are
\be\label {3x}\tilde E_{3}(x_{3}, t) = \int_{\R\times \R} d x_{1} dx_{2} \tilde e(\mathbf x, t) \qquad \tilde S_{3}(x_{3}, t) = \int_{\R\times \R} d x_{1} dx_{2} \tilde s(\mathbf x, t) ,\ee
$$ \tilde e(\bx, t) = {1\over 2} |\bu(\bx, t)|^{2}, \qquad \tilde s(\bx, t) =  \left |\nabla \bu(\bx, t)\right |^{2}.$$
 We  also consider the transverse marginals $E_{j}(k_{j},t)$, $\tilde E_{j}(x_{j},t)$, $S_{j}(k_{j},t)$, $\tilde S_{j}(x_{j},t)$, $j=1,2$. which are defined in an obvious way.
\par\smallskip
 
For the initial data (\ref{iniziali}),  if the constant $C$ is large enough, beyond some critical value corresponding to the initial energy to $E_{0} > 25,000$,   the solution  blows up after a time of the order $10^{-4}$ time units.   
 However, as in our screening  we never went beyond a time of the order $10^{-2}$, it may well be that some of the cases with lower initial energy do in fact blow up at a later time. The question whether the critical value is real remains open.

 \par \smallskip
The qualitative behavior of the solution    does not change  when we increase the constant $C$ in (\ref{iniziali}) beyond the (apparent) critical value. Therefore we only report results  with initial data (\ref{iniziali})  where $C$ corresponds to the initial  energy is $E_{0}= 200 \times (2\pi)^{3}\approx 49,500$. That is, we report results of only two solutions, of type $I$ and $II$.\par\smallskip
The rapid growth of the energy and enstrophy  takes place in a very short time with respect to the total run time, as shown in Fig. 1 for solutions of type $I$ and Fig. 2 for solutions of type $II$. Observe also  that the  enstrophy starts growing  earlier than the energy, and  the critical time is much smaller for the solution of type $I$, although the initial energy and enstrophy are the same. 

\begin{figure}[H]  
\centerline {
\includegraphics[width=3.in]{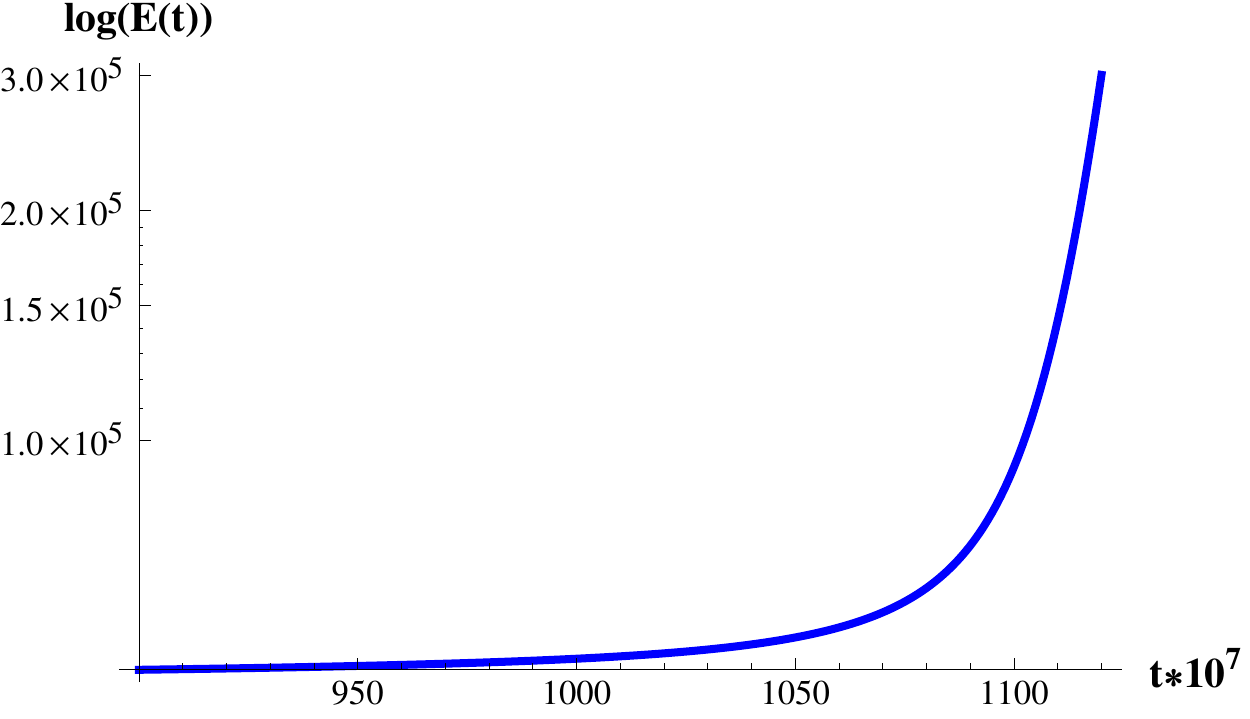}\hfill \includegraphics[width=3.in]{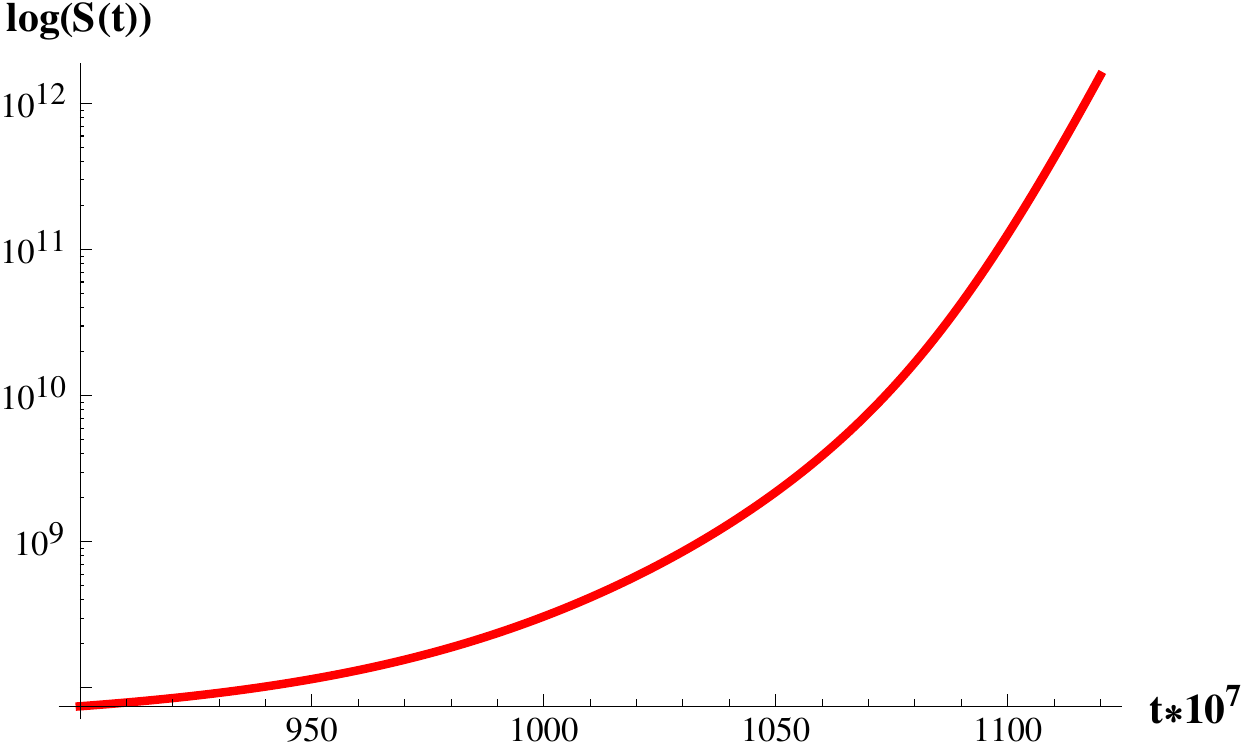}
}  
\caption{\it Log-plot of the total energy $E(t)$ (left) and the total enstrophy $S(t)$ (right) for the solution of type $I$ vs magnified time $t\times 10^{7}$. The vertical scale is logarithmic.  Longitudinal simulation range $k_{3} \in [-19, 2528]$.} \label{Fig.1a}\end{figure}
\par\smallskip 

\begin{figure}[H]  
\centerline {
\includegraphics[width=3.in]{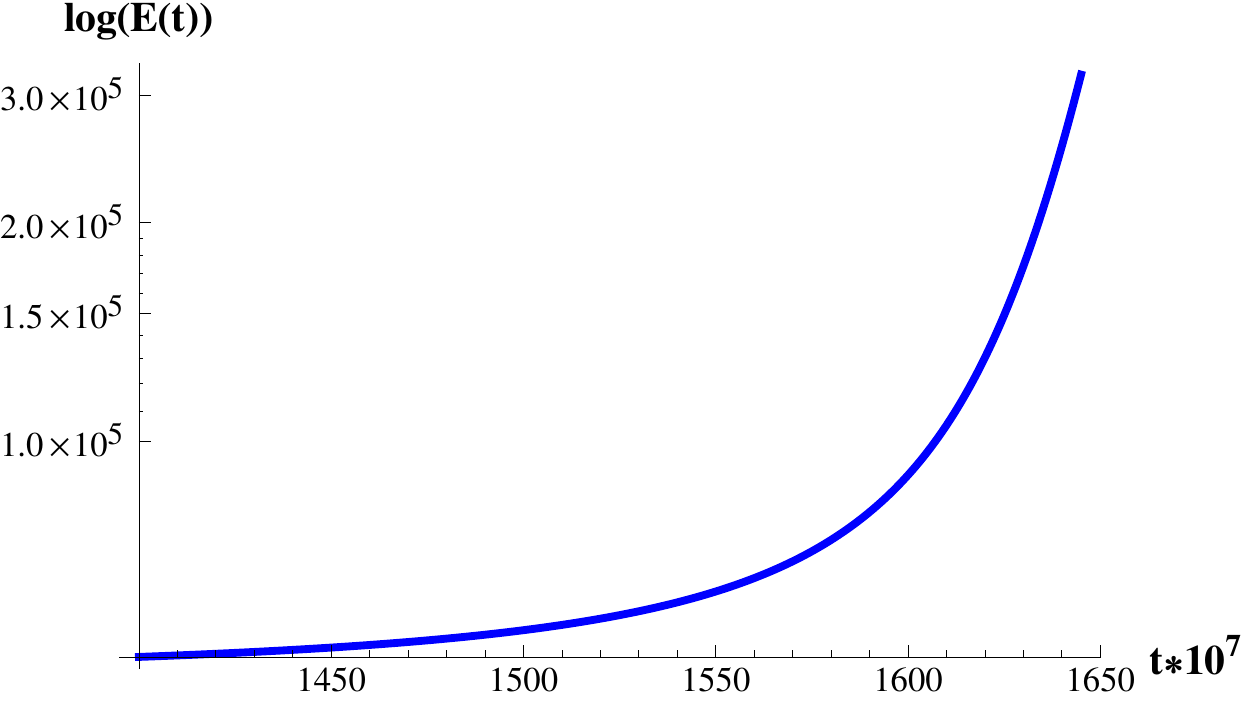}\hfill \includegraphics[width=3.in]{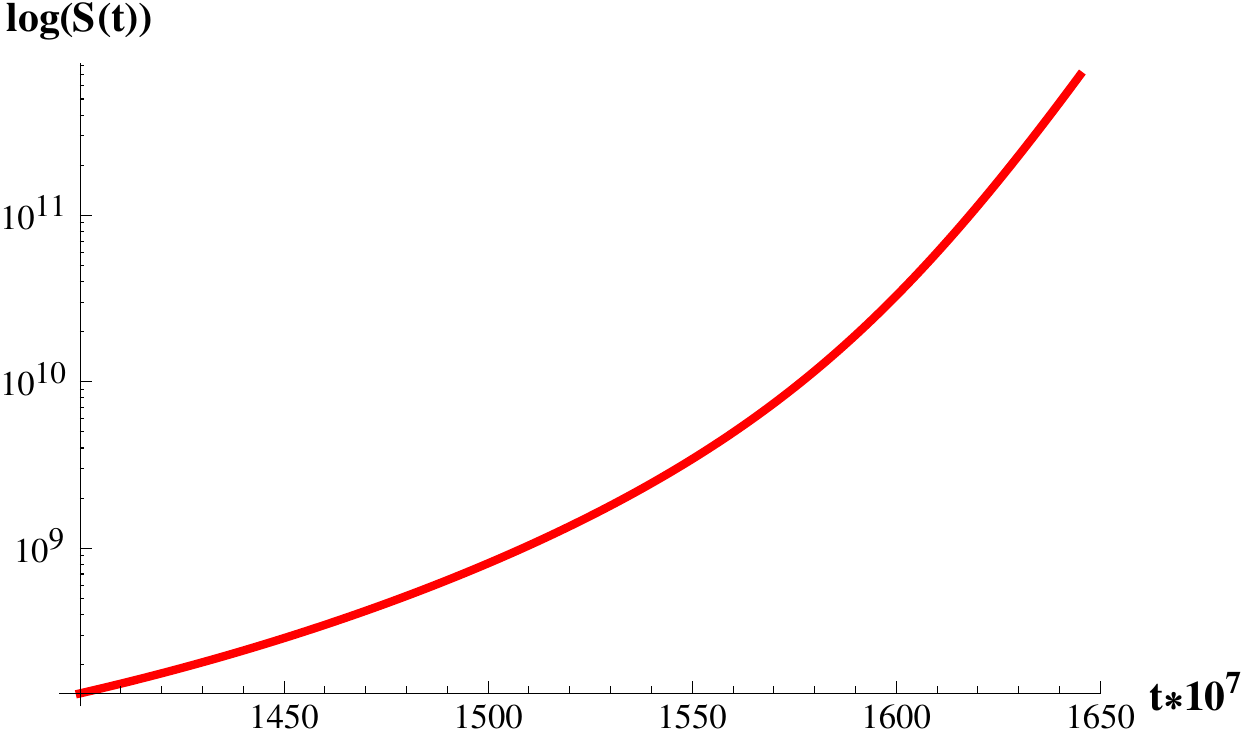}
}  
\caption{\it Log-plot of the total energy $E(t)$ (left) and the total enstrophy $S(t)$ (right)   for the solution of type $II$ vs magnified time $t\times 10^{7}$.   Longitudinal simulation range $k_{3} \in [-19, 2528]$.} \label{Fig.1b}\end{figure}
\par
The concentration of the support of the solutions around the $k_{3}$-axis and the role of the fixed point $\mathbf H^{\zer}$ are already visible at the beginning of the blow-up and for relatively small values of $k_{3}$. As shown in Fig. 3 for the solution of type $II$, the direction of the vector field on a plane orthogonal to the $k_{3}$-axis  is  approximately radial, and the absolute value $|\bv(\bk, t)|$  falls off sharply as we move away from the $k_{3}$-axis. 

 \begin{figure}[H]  
\centerline {
\includegraphics[width=3.5in]{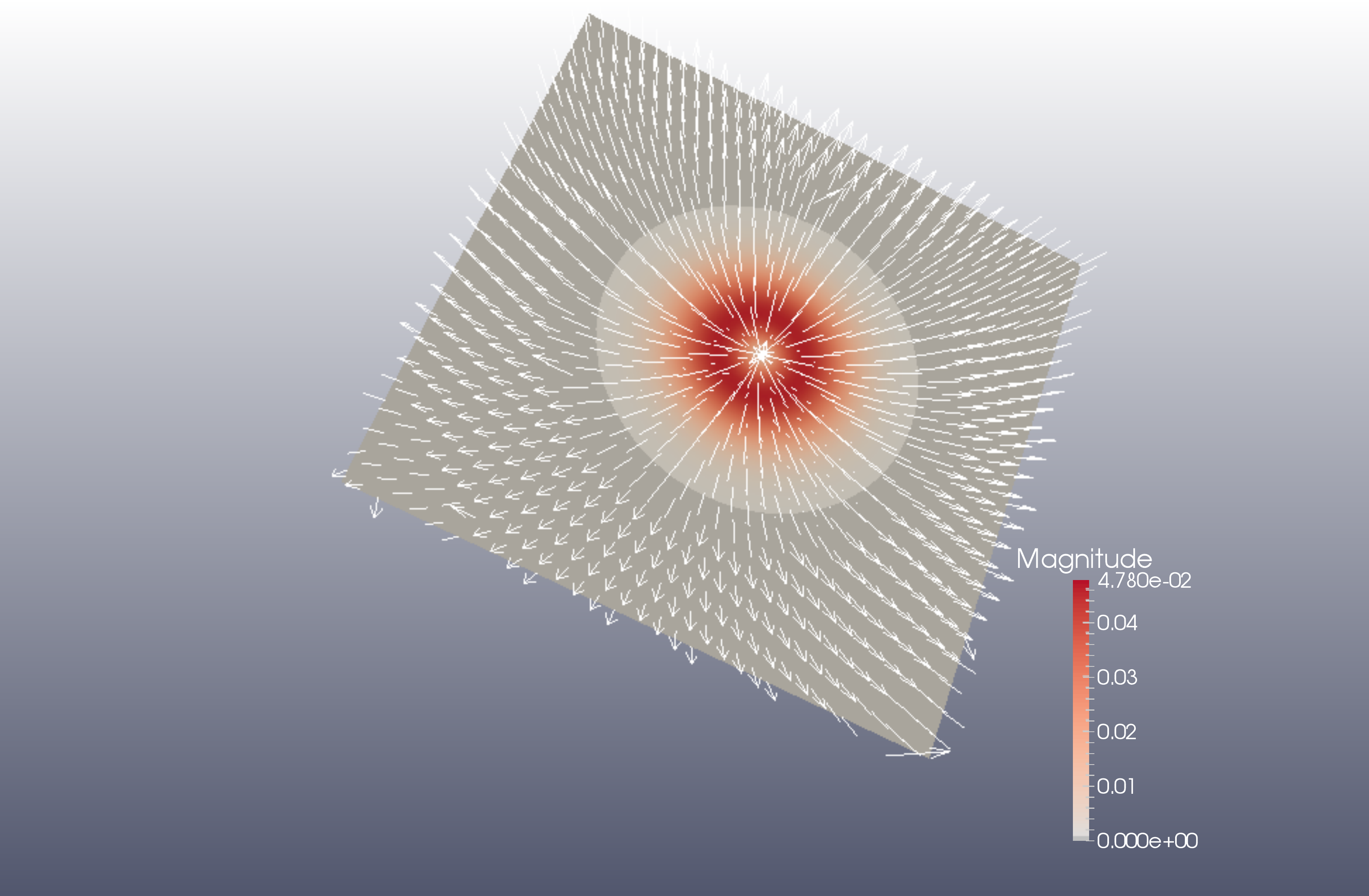}
}  
\caption{\it The arrows indicate the direction of the field $\bv(\bk,t)$  at the points of a uniform sublattice of a regular lattice of step $1$, for the solution of type $II$ on a square section of the plane $k_{3}=100$ with sides parallel to the axes and of length $100$, for $t=1521 \times 10^{-7}$. Longitudinal simulation range $k_{3} \in [-19, 2528]$.  The magnitude indicated in the figure refers to $|\bv(\bk, t)|$ and the grey external region indicates  values not exceeding $10^{-6}$.}\label{Fig.2}\end{figure}
\par\smallskip

The different behavior of the solutions of type $I$ and $II$ is illustrated in Fig. 4, which shows a local plot of the marginal enstrophy density $S_{3}(k_{3},t)$. For the solution of type $II$  we see sharp peaks, approximately at the points $k_{3} = p a$, $p=1,2,\ldots$, corresponding to the centers of the Gaussian factors in the expansion  (\ref{mainp}), with  zeroes between them, due to interference of neighboring peaks with different sign. The solution of type $I$ shows instead very mild peaks, and only for low values of $p$. \par As we shall see in the next section, the different behavior in $\bk$-space implies a different location of the singular point   in physical $\bx$- space.

\begin{figure}[H]  
\centerline {
\includegraphics[width=3.in]{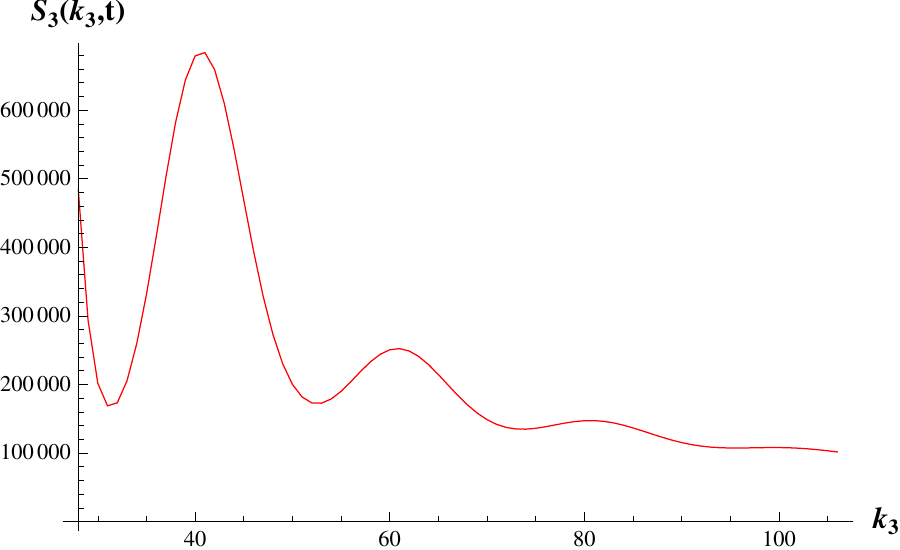}\hfill \includegraphics[width=3.in]{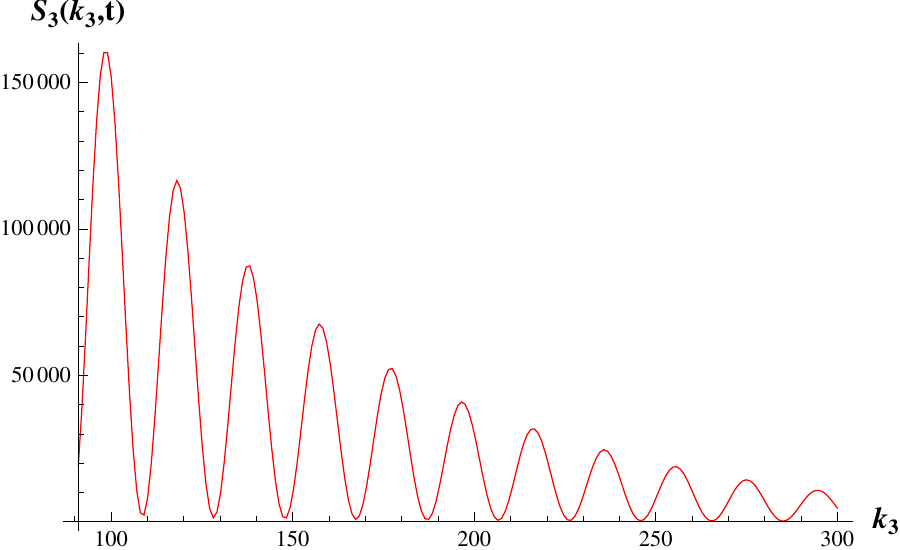}
}  
\caption{\it Plot of the enstrophy marginal density  $S_{3}(k_{3},t)$ for the solution of type $I$ (left) and $II$ (right)  at the beginning of the blow-up.  The time is $t= 900 \times 10^{-7}$  for type $I$ and  $t= 1125 \times 10^{-7}$ for type $II$.  The  zeroes for the type $II$ solution are approximately  periodic with period $a= 20$. Longitudinal simulation range $k_{3} \in [-19, 2528]$.} \label{Fig.5}\end{figure}

The main difficulty in following the blow-up by computer simulations is that, as we get close to the critical time, the essential support of the solutions moves away along the $k_{3}$-axis. The divergence of the energy could not be followed in a proper way, because even in the initial times of the blow-up its growth is due to modes with $k_{3}$ exceeding the maximal value available to us $L= 3025$. 
\par\smallskip
 The divergence of the enstrophy is easier to follow, because, as we mentioned above, it starts growing much earlier than the energy, when the range of the values of $k_{3}$ which contribute to the growth is  contained within the simulation region $[-19, L]$, for $L\geq 2028$. \par\smallskip
 Fig. 5 shows the behavior in time of the enstrophy marginal density $S_{3}(k_{3},t)$ along the $k_{3}$-axis. It can be seen that in the range of times under consideration there is a significant increase of the enstrophy due to the modes with $k_{3}$ within the simulation range. 
The transversal distribution of the enstrophy given in Fig. 6, for type $II$ and the  same times as in Fig. 5,   shows that the  relevant support of the enstrophy   grows very slowly in the transversal direction and is well contained inside the computation region  $[-127, 127]\times [-127, 127]$. 

\par\smallskip
 Fig. 5 also shows that, as we approach the critical time, the main support of $S_{3}(k_{3},t)$ moves away to the high $k_{3}$ region,  while the function for low $k_{3}$ values, to the left of the growing maxima,   changes very little. This behavior is in accordance with  the  point-wise convergence in $\bk$-space, which was deduced in Section 2.\par\smallskip
   The rate of convergence of the solution  for low $k_{3}$ values is given for type $II$ in Fig. 7, where a transversal component of $\bv(\bk, t)$ is reported as a function of  $k_{3}$,    for $k_{1}, k_{2}$   fixed,  and several values of $t$.  Observe that the behavior is a kind of  damped oscillation with period $2a$, and zeroes at the points $k_{3}\approx {1\over 2} (2j+1)\; a$, $j=1, \ldots$. The other transversal component behaves in the same way. 
\par \smallskip 

\begin{figure}[H]  
\centerline {
\includegraphics[width=3.in]{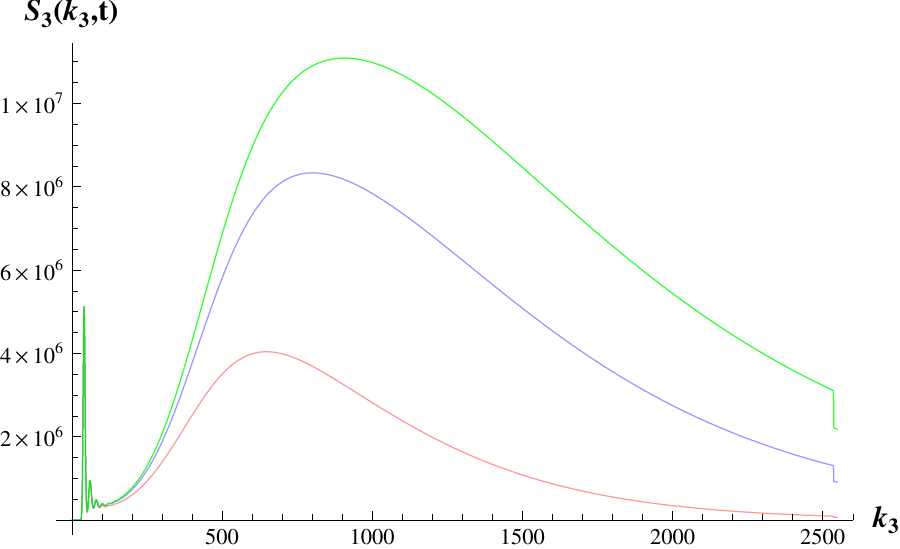}\hfill \includegraphics[width=3.in]{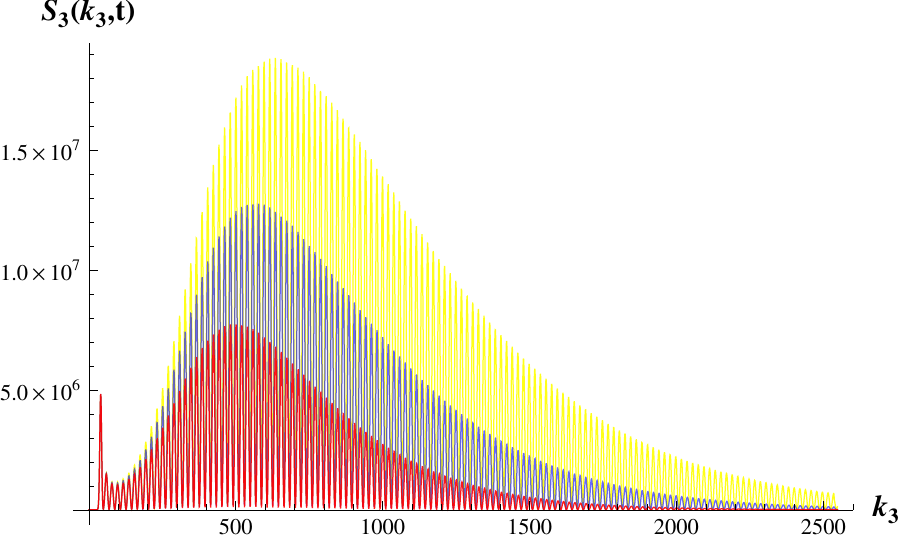}
}  
\caption{\it Plot of the   marginal enstrophy density $S_{3}(k_{3},t)$ on the whole simulation range  $-19 \leq k_{3}\leq 2528$, for type $I$ (left) at $t \cdot 10^{7} = 1060,  1075, 1080$, and for  type $II$ (right) at $t \cdot 10^{7} = 1521,  1544, 1560$.   Longitudinal simulation range $k_{3} \in [-19, 2528]$.} \label{Fig.5}\end{figure}

    \begin{figure}[H]  
\centerline {
\includegraphics[width=3.5in]{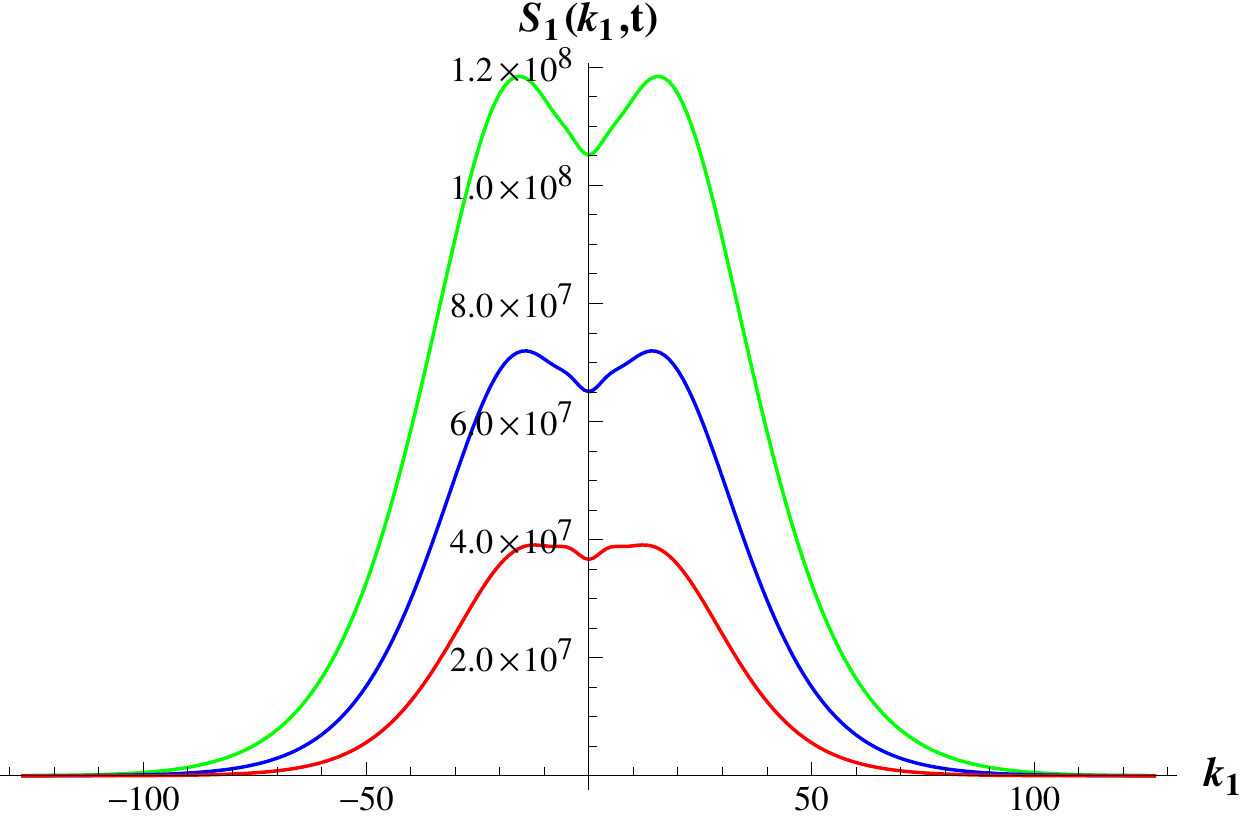}
}  
\caption{\it Solution of type $II$: plot of the marginal enstrophy density $S_{1}(k_{1},t)$  at $t \cdot 10^{7} = 1521,  1544. 1560$  Longitudinal simulation range $k_{3} \in [-19, 2528]$.}\label{Fig.10}\end{figure}

 \begin{figure}[H]  
\centerline {
\includegraphics[width=3.5in]{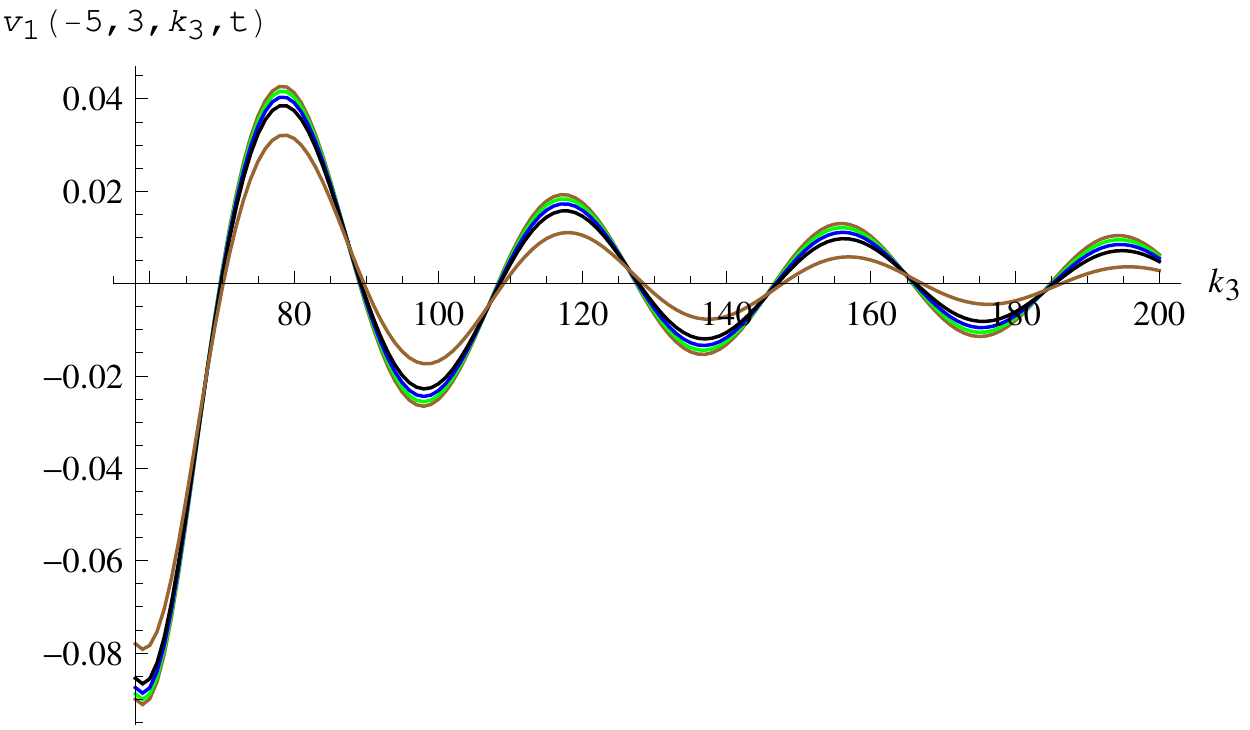}
}  
\caption{\it Solution of type $II$: plot of the transversal component of the velocity $\bv_{1}(\bk, t)$ for $k_{1}, k_{2}$ fixed vs  $k_{3}$, at the times   $t \times 10^{7}=1342, 1500, 1544, 1574, 1600$.   For identification, observe that as time grows the oscillation amplitudes increase. Longitudinal simulation range $k_{3} \in [-19, 3028]$.}\label{Fig.4}\end{figure}

The problem of  estimating  the critical time $\tau$  is rather challenging because the divergence is due to the large $k_{3}$ modes, which act in a different way on the enstrophy and energy. The most reliable way of estimating $\tau$ is that based on formula (\ref{mainp}), which indicates that, except for power-law corrections, the absolute value of the solution $|\bv_{A(\tau)}(\bk,t)|$ for $\bk_{\perp}$ fixed, falls off exponentially fast in $k_{3}$ with a rate proportional to $(\tau-t)$.  \par \smallskip

The same property extends to the marginal 
 energy density $E_{3}(k_{3}, t)$, and indeed, as shown by Fig. 8 for type $II$,  it is verified for $k_{3}\geq 400$ already at times relatively far from the critical time. The points align on a straight line with great accuracy, and the slope decreases with time. (The dots represent the local maxima, as the marginal density $E_{3}$, in analogy with the  enstrophy density $S_{3}$ shown in Figg. 4, is wildly oscillating.) Moreover the exponential decay rate is remarkably stable with respect to the longitudinal simulation range for $L\geq 2028$, for solutions of both types. 

 \begin{figure}[H]  
\centerline {
\includegraphics[width=3.5in]{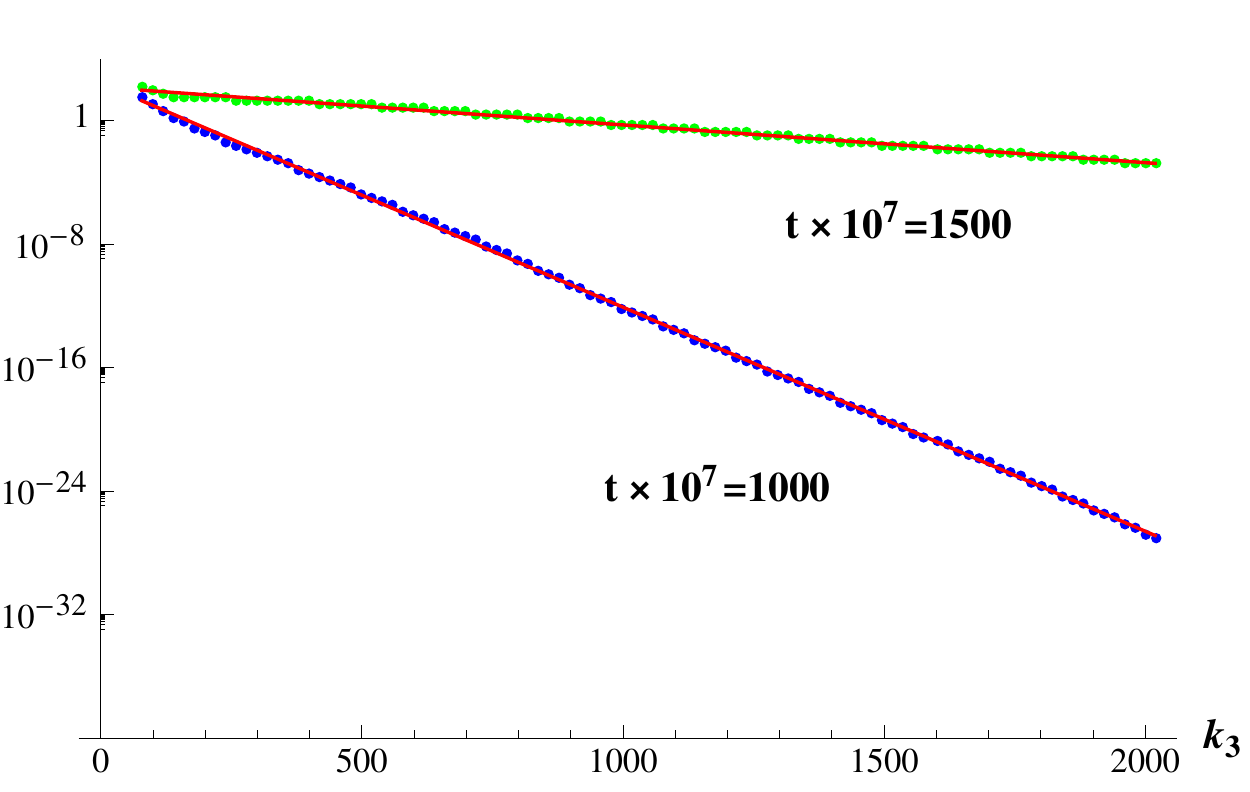}
}  
\caption{\it  Type II: plot of $\log (E_{3}(k_{3},t))$, where $E_{3}$ is the marginal energy density alon the $k_{3}$-axis for $k_{3}\geq 400$ at two different times. The dots represent the local maxima   of the oscillations of $E_{3}(k_{3},t)$. Longitudinal simulation range $k_{3} \in [-19, 2028]$.}\label{Fig.11}\end{figure}
\par\smallskip

 If we now plot the exponential decay rate vs time, as shown on Fig.9 for both types, we get an estimate of the critical time by looking at the intercept with the horizontal axis.  The estimates are $\tau \approx 1110 \times 10^{-7}$ for type $I$ and $\tau \approx 1630 \times 10^{-7}$ for type $II$.  Observe that all estimates are to be considered as  overestimates, because the simulations do not take into account the contribution of the modes outside the simulation region, which enhances the divergence.

\begin{figure}[H]  
\centerline {
\includegraphics[width=3.in]{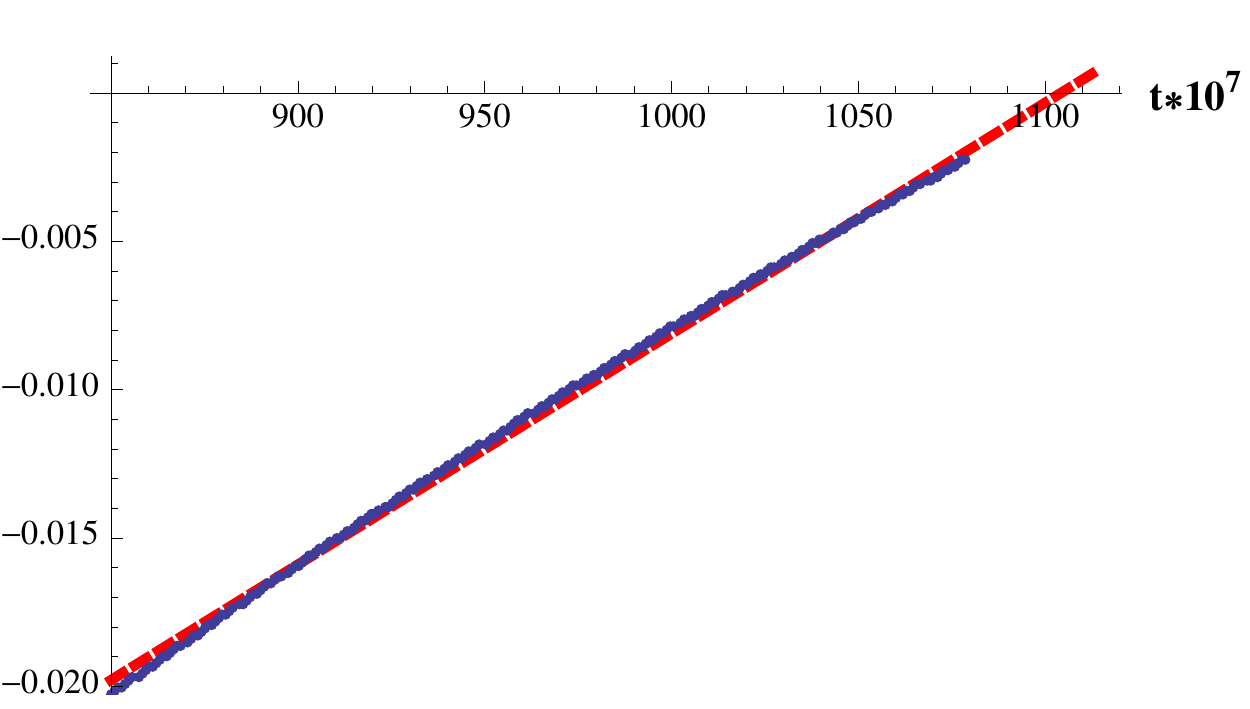}\hfill \includegraphics[width=3.in]{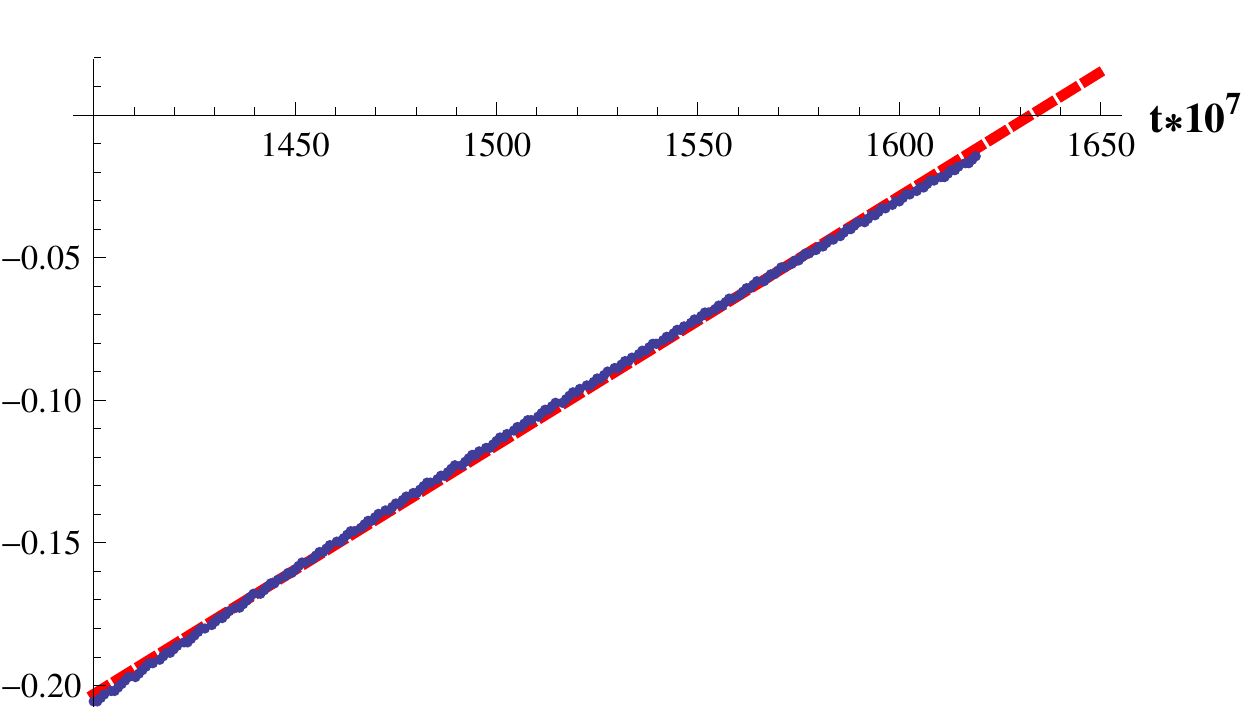}
}   
\caption{\it  Exponential decay rate (slope of the log-plots as in Fig. 8) for the marginal density $E_{3}(k_{3},t)$, taken for $k_{3}\geq 400$ vs magnified time $t\times 10^{7}$, for type $I$ (left) and type $II$ (right), with  linear regression (dashed line). Longitudinal simulation range $k_{3} \in [-19, 2528]$.}\label{Fig.12}\end{figure}
\par\smallskip

We conclude this section with a  check of the rates of divergence (\ref{divergenza}) predicted in Section 2. We restrict our considerations  to the divergence of the total enstrophy, because,  as we explained above, the data on the growth of the energy that we could collect are not good enough.
 In Fig.10 we report  a plot of $\log S(t)$ versus $\log {1\over \tau_{*}- 10^{7} \;t}$, where  $\tau_{*}$ is the estimated critical time in magnified units ($\tau_{*}= 1110$ for type $I$, and $\tau_{*}= 1630$ for type $II$)  in the range of times which are sufficiently large and within the reliable range.  The results should be compared with the prediction of (\ref{divergenza}), which is a power $\al =3$ for type $I$ and $\al = 2.5$ for type $II$.  Taking into account that that the slopes in Fig. 10 are very sensitive to the  value of the predicted critical time, we  conclude that the results, namely $\al \approx 3,12$ for type $I$ and $\al \approx 2,6$ for type $II$, are compatible with the predictions in (\ref{divergenza}).  \par\smallskip
More computer resources are needed in order to get a deeper understanding of the role of the high $k_{3}$ modes   on the divergence of the total energy and the total enstrophy as $t\uparrow \tau$.

\begin{figure}[H]  
\centerline {
\includegraphics[width=3.in]{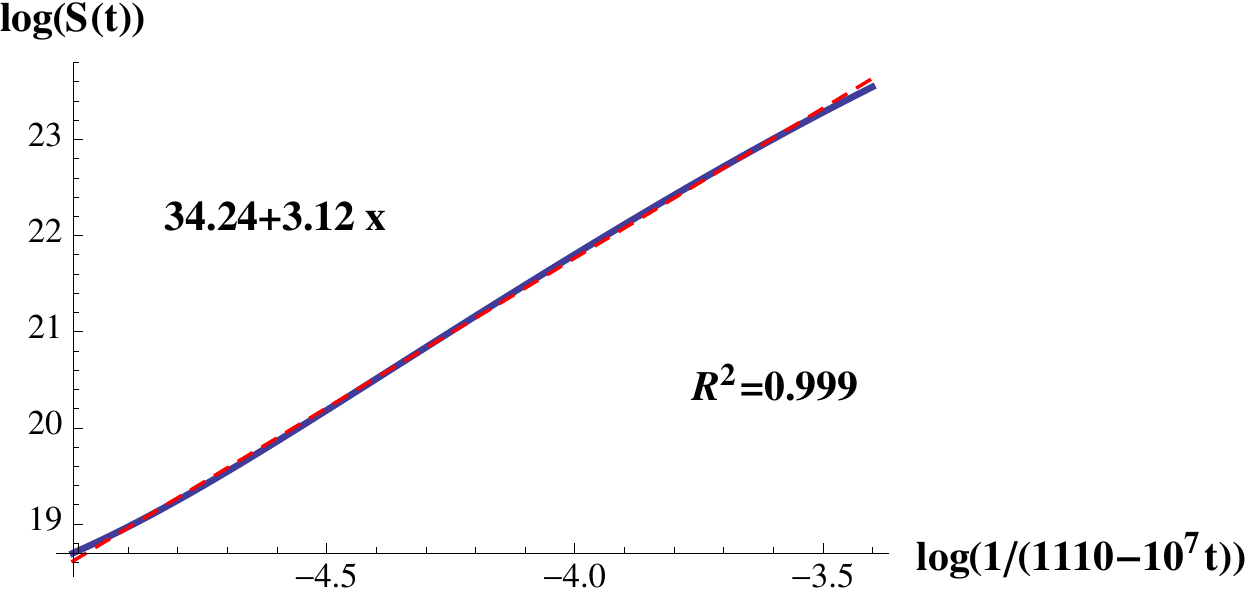}\hfill \includegraphics[width=3.in]{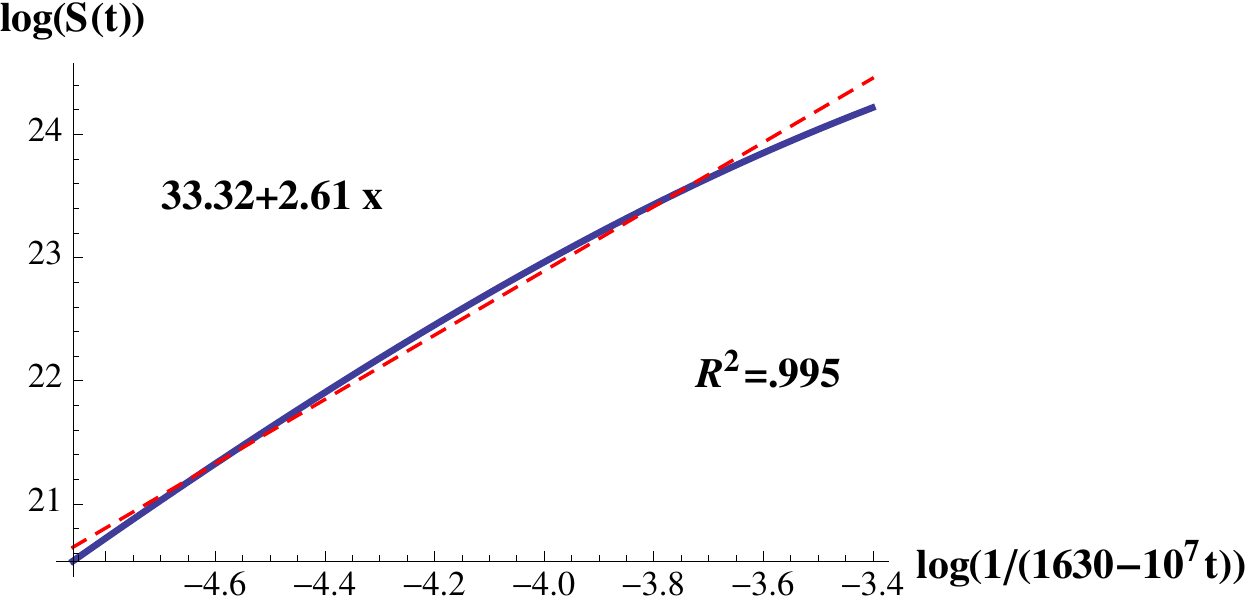}
}   
\caption{\it  Log-plot of  the total enstrophy $S(t)$ vs $\log {1\over \tau_{*}-t}$, at times near the blow-up, for type $I$ (left) and  type $II$ (right), with  linear regression (dashed line, with regression formula above the line). Longitudinal simulation range $k_{3} \in [-19, 2528]$.}\label{Fig.12}\end{figure}
\par\smallskip

 \par\smallskip

 \section {Computer simulations: the Li-Sinai solutions in $\mathbf x$-space}
 \label{S5}

In $\bx$-space the main support of the solution    is  contained within a small volume around the origin, and as $t\uparrow \tau$,  it concentrates in sharp ``spikes'', which become singular in the limit.  \par\smallskip
 As shown in Fig. 7, the solution    of type $II$  oscillates in $k_{3}$ with a period $T\approx 2a$ (with $a=20$), so that we expect large values of the velocity $\bu(\bx, t)$ around the points  $\mathbf x^{\zer}_{\pm}=
 (0, 0, \pm x_{3}^{\zer})$ with $x^{\zer}_{3}\approx 0.16 \approx {\pi\over a}$. 
 The behavior of the marginal  energy densities $\tilde E_{3}( x_{3},t)$ and  $\tilde E_{1}(x_{1}, t)$ for two different times given  in Fig. 11 and Fig. 12.
 It can be seen that  $\tilde E_{3}( x_{3},t)$ shows two spikes at the points $\pm x_{3}^{\zer}$, while  the transversal marginal $\tilde E_{1}(x_{1}, t)$ has a single spike at the origin. 
 Both spikes increase indefinitely  as $t\uparrow \tau$, while everywhere else both marginals seem to converge to a finite limit.\par\smallskip
  A possible conclusion, which however would require further investigation, is that   for all $\bx \neq \mathbf x^{\zer}_{\pm}$ the function $\bu(\bx, t)$ itself also converges. A result of this kind was in fact proved for the complex Burgers equations (see \cite{LiSi10} and \cite{BFM12} for computer simulations).

 \par\smallskip

The marginal $\tilde E_{3}(x_{3}, t)$ for the solution of type $I$, as  given in Fig. 13,  also for two different times, differs from the previous one, in that it has a single spike, suggesting that the only singularity of $\bu(\bx,t)$ as $t\uparrow \tau$ is at the origin. Here also the marginals indicate divergence at $\bx=0$ and convergence elsewhere.

 \par \smallskip 
   
    An important role   with respect to possible singularities of the NS equations is played by the vorticity stretching  vector
     $\mathbf w(\bx, t) =\omega(\bx, t) \cdot \nabla \mathbf u(\mathbf x, t)$, 
 where $ \omega(\bx,t)  = \nabla \times \bu(\bx,t)$ is the vorticity (see, e.g, \cite{RuG04}).  In Fig. 14 we show, for the solution of type $II$, a joint logarithmic plot of the enstrophy marginal $\tilde S_{3}(x_{3},t)$ and of the corresponding marginal for vorticity stretching
 $$W_{3}(x_{3},t) = \int_{\R\times \R} dx_{1}dx_{2}    \left |  \omega(\bx, t) \cdot \nabla \mathbf u(\mathbf x, t)\right |^{2},$$
    ($W_{3}$, for dimensional homogeneity, is divided by $E_{0}$). Observe that the spikes of the enstrophy marginal are more enhanced  than  for the energy, and those of $W_{3}$ even more so. The transverse marginals $\tilde S_{1}(x_{1},t)$, and the corresponding one for the vorticity stretching have a spike of similar magnitude at the origin.

 \begin{figure}[H]  
\centerline {
\includegraphics[width=3.5in]{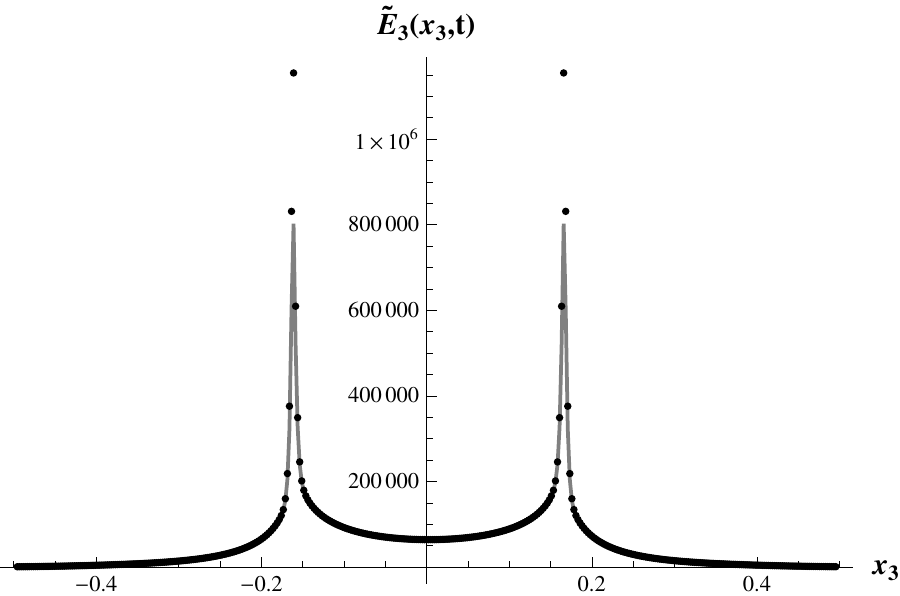}
}  
 \caption{\it Type $II$: plot of the marginal  energy density $\tilde E_{3}(x_{3},t)$ at $t \cdot 10^{7} = 1521$ (continuous line) and $t\times 10^{7}= 1544$ (dotted line). Longitudinal simulation range $k_{3} \in [-19, 2528]$}\label{Fig.15}\end{figure}
 
 \begin{figure}[H]  
\centerline {
\includegraphics[width=3.5in]{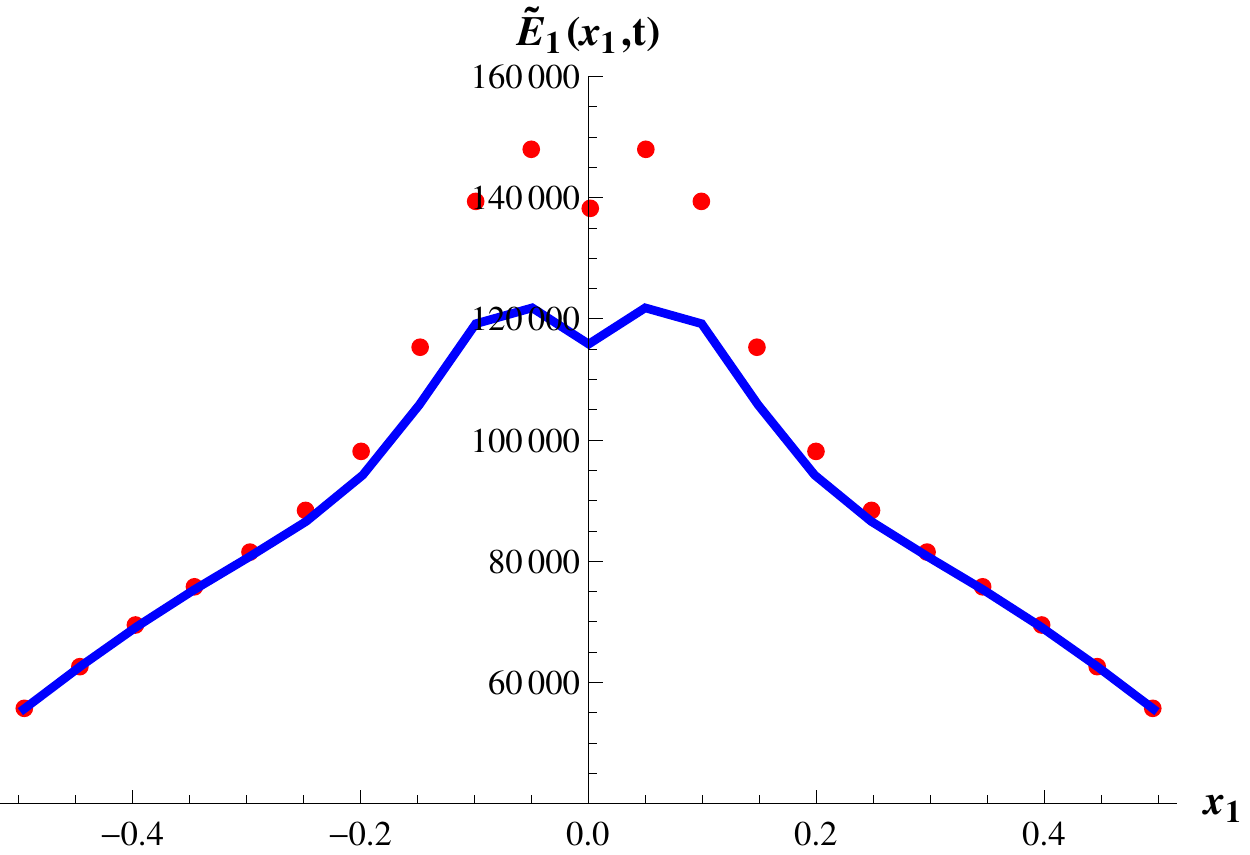}
}  
 \caption{\it  Type $II$: plot of the marginal  energy density $\tilde E_{1}(x_{1},t)$ at $t \cdot 10^{7} = 1521$ (continuous line),  and $t \cdot 10^{7} =1544$ (dotted line). Longitudinal simulation range $k_{3} \in [-19, 2528] $.}\label{Fig.14} \end{figure}

 \begin{figure}[H]  
\centerline {
\includegraphics[width=3.5in]{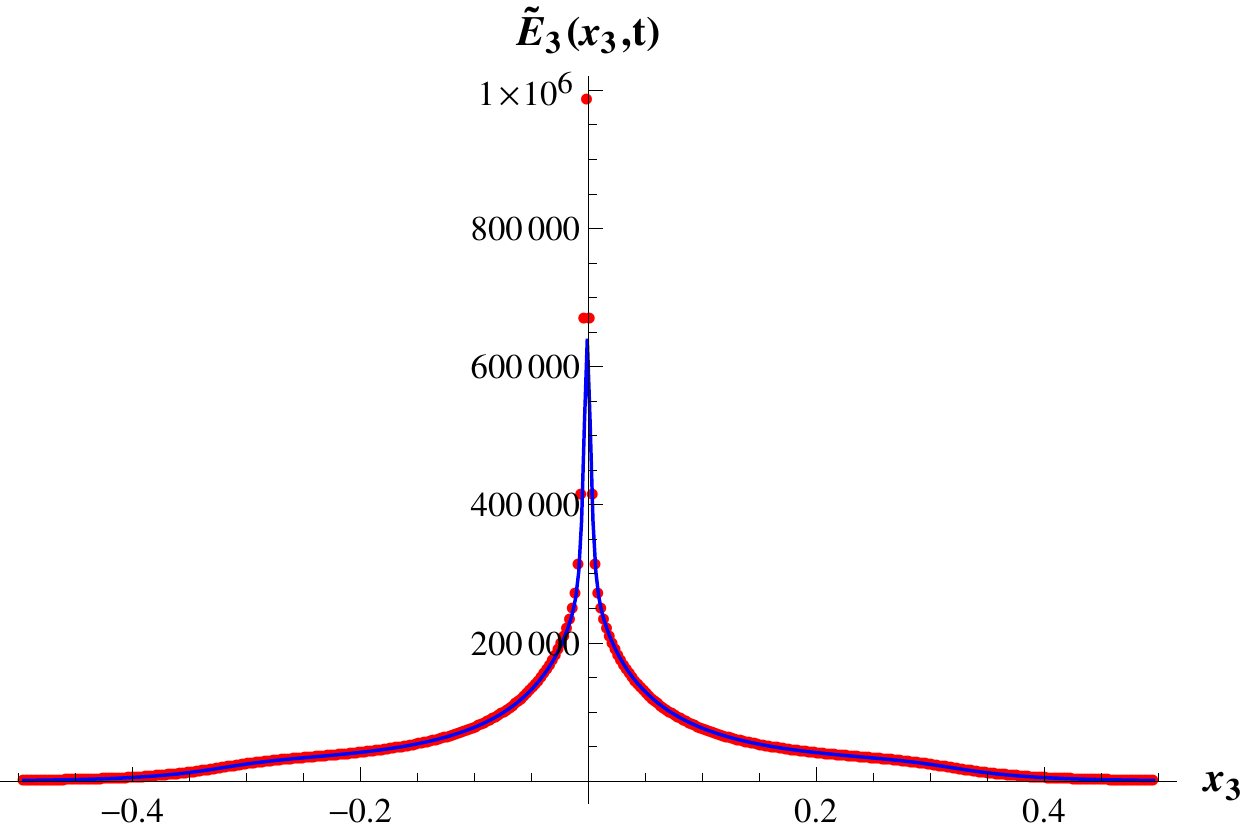}
}  
 \caption{\it Type $I$: plot of the marginal  energy density  $\tilde E_{3}(x_{3},t)$ at $t \cdot 10^{7} = 1021$ (dotted line) and $t\times 10^{7}= 1044$ (continuous line). Longitudinal simulation range $k_{3} \in [-19, 2528]$.}\label{Fig.15}\end{figure}

 \begin{figure}[H]  
\centerline {
\includegraphics[width=3.5in]{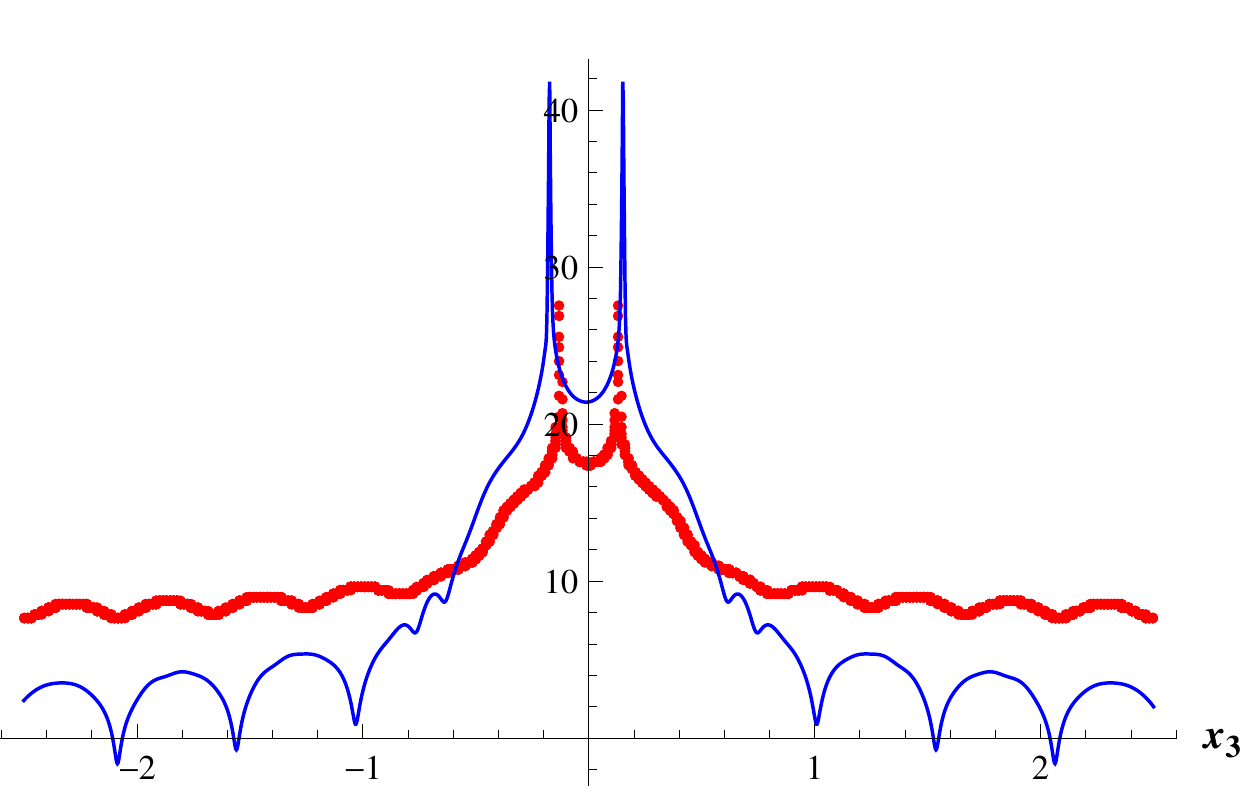}
}  
 \caption{\it Type $II$: Log-plot of the marginal enstrophy density  $\tilde S_{3}(x_{3},t)$ (dotted line) and of the vortex stretching density  $ W_{3}(x_{3},t)$ (continuous line)  at $t \cdot 10^{7} = 1544$. Longitudinal simulation range $k_{3} \in [-19, 2528]$.}\label{Fig.24}\end{figure}

\section {Some information on the computer simulations·}
\label {S6}

As we said above, we simulate the integral equation (\ref{kequation}) in Fourier $\bk$-space, and it is of great help the fact that the support of the solutions in $\bk$-space is concentrated  along the $k_{3}$-axis. \par\smallskip
The discretization   is implemented by a regular mesh of points containing the origin, and such that it contains the region where the solution is significantly non-zero.  For all simulations reported in this paper the mesh in $\bk$-space is taken uniform with  step $1$, and is a set of the type  $ R = [-127, 127] \times [-127, 127] \times [-19, L] \subset \Z^{3}$ (the brackets $[\ldots ]$ denote  intervals in  $\Z$), with $L= 2028, 2528, 3028$.  Control simulations with a refined mesh were performed to check stability, which showed that
the results  are remarkably stable with respect to refinements of the mesh, in accordance with the fact  that   the solution $\bu(\bx,t)$ in $\bx$-space is essentially concentrated in a small region around the origin (see \S 5).  \par\smallskip

However, as we approach the critical time, the simulations are  very sensitive to the time step,  and to the strip length $L$.    We always checked that the time step $\delta_{t}= 10^{-7}$ and the given value of $L$ are such as to ensure stability in the range of times under consideration.

     \par\smallskip
 
 The discrete computation
 is obtained by the so-called Nystr\"om method with respect to the
$\bk$ variables and a predictor-corrector scheme with respect to the time variable,
which is based on the Euler method and the Trapezoidal method.
\par\smallskip
The predictor-corrector scheme iterates a recursive procedure to compute successive approximations $\mathbf V^{(j)}(\bk, n \delta_{t})$, $j=1,2,\ldots$,  of the function $\bv(\bk, n\delta_{t})$, until a convergence criterion is satisfied: $|\mathbf V^{(j+1)}(\bk, n\delta_{t}) -\mathbf V^{(j+1)}(\bk, n\delta_{t}) |\leq {\rm tol}$, where the tolerance ${\rm tol}$ is   set at $10^{-8}$. 
\par\smallskip
The procedure is computationally challenging, and in
fact for each $\bk$ in the mesh it requires the evaluation of a three-dimensional integral. The integral is 
 a convolution,  and by using the fast Fourier transform (FFT) we can  reduce
the computational cost. Note that, as we are simulating the NS equations in $\R^{3}$,  the convolution is not  periodic, and has to be
 implemented on a computational grid which is doubled in size.
\par\smallskip

The accuracy of the approximated solution corresponding to the chosen initial data is evaluated on an experimental basis by
comparing the results obtained for different discretization parameters.

\par\smallskip

  Our computer simulations were performed at
   CINECA of Bologna (Italy) on  the  FERMI Supercomputer (Model: IBM-BlueGene/Q; Architecture: $10$ BGQ Frame; Processor
Type: IBM PowerA2, $1.6$ GHz; Computing Cores $163840$; Computing Nodes
$10240$; RAM: 1GByt/core)   
   \par\smallskip
The computation method was implemented in Fortran 90 (IBM Fortran compiler)
with MPI library for parallel computations, and 2Decomp\&FFT for the
parallel computation of the fast Fourier transform.

 \section {Concluding Remarks}
 \label{S7}
 \par\smallskip
 
We deduced in Section 2 some important consequences of the work of Li and Sinai \cite{LiSi08} which clarify the behavior of the solutions near the blow-up.  With the help of computer simulation it was then  possible to check the predictions and  to estimate the critical time. Moreover,   what is perhaps more relevant, the computer simulations give a detailed picture of the behavior in $\bx$-space, indicating important properties of the solutions, such as the point-wise  convergence as $t\uparrow \tau$  in $\bx$-space, except for the singular points, which are not so far predicted by the theory, but can hopefully be proved rigorously in the near future.   \par\smallskip

 We would like to remark that simulations of the solutions of the 3-d NS equations are usually computationally onerous and sometimes unreliable, especially for flows with large values of the enstrophy (see \cite{Hou09} for a review).     It is a remarkable fact that the singular complex  solutions proposed by Li and Sinai, due to their simple structure in $\bk$-space, are relatively easy to follow  by computer simulations on the supercomputers of the last generation.  
 
 \par\smallskip
 It is of particular importance in this respect that the solution can be represented as a power series (\ref{serie}), where the parameter $A$ governs the blow-up time. A great help also comes from the stability of the computation with respect to the discretization step in $\bk$-space, which is due to the confinement of the energy in a small region of $\bx$-space.   In fact,  in the original  paper  \cite{LiSi08} the infinite extension of the domain in $\bx$-space does not seem to be  essential, except for the absence of  boundary conditions, and the proofs can possibly  be adapted to the periodic case on the torus $T^{3}$.

  \par\smallskip

  The general picture that comes out is that of a   motion in which the ``fluid'' moves very fast in a small region around the origin for solutions of type $I$ or around
   two symmetric points close to the origin for solutions of type $II$,  along flow lines with high curvature, as indicated by the behavior of the vorticity and the vorticity stretching described in \S 5. \par \smallskip

 By antisymmetrizing the initial data one gets real-valued solutions which share some basic  properties of the complex solutions, such as the restriction of the support to a thin cone in $k$ space. The study  of   such solutions, which for some interval of time show strong similarities to   solutions of type $II$, is in progress.

\section {Acknowledgements} 
\label{S8}
We thank Prof. Ya. G. Sinai and Dr. D. Li for their constant interest on our work, and for many discussions and suggestions.   We acknowledge the CINECA award under the ISCRA initiative IsB10\_3DNS (2014) and IsC23\_3DNS (2015),  for the availability of high performance computing resources and support.

\par\bigskip \noindent 
{\bf Funding}  To S.F.: Ministero dell'Istruzione, dell'Universit\`a e della Ricerca, COFIN Mat/07 2012.

\bibliographystyle{imamat}

       \end{document}